\documentclass[journal]{IEEEtran}
\usepackage{amsmath,amsfonts}
\usepackage{algorithmic}
\usepackage{algorithm}
\usepackage{array}
\usepackage[caption=false,font=footnotesize,labelfont=rm,textfont=rm]{subfig}
\usepackage{textcomp}
\usepackage{stfloats}
\usepackage{url}
\usepackage{verbatim}
\usepackage{graphicx}
\usepackage{cite}
\usepackage{amsthm}
\usepackage{amssymb}
\usepackage{epsfig}
\usepackage{graphicx}
\usepackage{balance}
\usepackage{epstopdf}
\usepackage{multicol}
\usepackage{multirow}
\usepackage{threeparttable}
\usepackage{booktabs}
\usepackage{cite}
\usepackage{cases}
\usepackage{cuted}
\usepackage{xcolor}
\usepackage{threeparttable}
\usepackage{amsmath}
\allowdisplaybreaks[2]

\newcommand\relphantom[1]{\mathrel{\phantom{#1}}}
\usepackage{geometry}
\begin{document}

\title{Graph Chirp Signal and Graph Fractional Vertex--Frequency Energy Distribution}

\author{Manjun~Cui, Zhichao~Zhang,~\IEEEmembership{Member,~IEEE}, and Wei~Yao
	\thanks{This work was supported in part by the Open Foundation of Hubei Key Laboratory of Applied Mathematics (Hubei University) under Grant HBAM202404; and in part by the Foundation of Key Laboratory of System Control and Information Processing, Ministry of Education under Grant Scip20240121. \emph{(Corresponding author: Zhichao~Zhang.)}}
	\thanks{Manjun~Cui and Wei~Yao are with the School of Mathematics and Statistics, Nanjing University of Information Science and Technology, Nanjing 210044, China (e-mail: cmj1109@163.com; yaowei@nuist.edu.cn).}
	\thanks{Zhichao~Zhang is with the School of Mathematics and Statistics, Nanjing University of Information Science and Technology, Nanjing 210044, China, with the Hubei Key Laboratory of Applied Mathematics, Hubei University, Wuhan 430062, China, and also with the Key Laboratory of System Control and Information Processing, Ministry of Education, Shanghai Jiao Tong University, Shanghai 200240, China (e-mail: zzc910731@163.com).}}

\markboth{IEEE TRANSACTIONS ON SIGNAL PROCESSING}
{Shell \MakeLowercase{\textit{et al.}}: Bare Demo of IEEEtran.cls for Journals}


\maketitle

\begin{abstract}
Graph signal processing (GSP) has emerged as a powerful framework for analyzing data on irregular domains. In recent years, many classical techniques in signal processing (SP) have been successfully extended to GSP. Among them, chirp signals play a crucial role in various SP applications. However, graph chirp signals have not been formally defined despite their importance. Here, we define graph chirp signals and establish a comprehensive theoretical framework for their analysis. We propose the graph fractional vertex--frequency energy distribution (GFED), which provides a powerful tool for processing and analyzing graph chirp signals. We introduce the general fractional graph distribution (GFGD), a generalized vertex--frequency distribution, and the reduced interference GFED, which can suppress cross-term interference and enhance signal clarity. Furthermore, we propose a novel method for detecting graph signals through GFED domain filtering, facilitating robust detection and analysis of graph chirp signals in noisy environments. Moreover, this method can be applied to real-world data for denoising more effective than some state-of-the-arts, further demonstrating its practical significance.
\end{abstract}


\section{Introduction}
\indent Graph signal processing (GSP) is an emerging field that analyzes and processes signals defined on  irregular graph structures \cite{Ortega08,ortega22introduction,Leus23}. It extends traditional signal processing (SP), which typically operates in Euclidean domains, to non--Euclidean domains, facilitating the analysis of data from networks such as social, sensor, and biological systems, where traditional SP may be inadequate \cite{Ortega08,Shuman13,Sandryhaila13,Sandryhaila14,Ljubisa20}.

\indent In SP, the fractional Fourier transform (FRFT) \cite{Almeida94,Oza01} is a generalization of the classical Fourier transform (FT) \cite{bochner1949fourier,nussbaumer1982fast} that facilitates a more flexible analysis of signals in different fractional frequency domains. The FRFT is an effective tool for analyzing and processing linear frequency modulation (LFM) signals, also known as chirp signals, which are characterized by their frequency variation over time \cite{Stank93,Djuric90,Almeida94,Cowell10,white2012performance}. {In GSP, the graph Fourier transform (GFT) \cite{Sandryhaila13,Shuman13,ortega22introduction,Gavili17} has been extended to graph fractional Fourier transform (GFRFT) \cite{Wang17,Alikasifoglu24} by introducing the fractional parameter $a$, facilitating a more sophisticated signal analysis in fractional graph spectral domains. However, the class of graph signals particularly suited for analysis using GFRFT remains unexplored in the current GSP literature. Motivated by the close connection between FRFT and chirp signals, we can define the graph chirp signals as natural counterparts in the graph domain. }

\indent Time--frequency analysis is a fundamental technique in SP \cite{Stankovic94,Coh95}.  Time--frequency distributions provide simultaneous representations of signals’ attributes in the combined time--frequency domain, enabling a more comprehensive analysis of signals with time-varying frequency content. {Among these, the Wigner-Ville distribution (WD) stands out as one of the most widely used methods \cite{Wig32, Sad21,Boashash87}, owing to its high resolution and capability to accurately capture instantaneous frequency and temporal dynamics. The relationship between WD and FRFT reflects a rotation of the time--frequency plane \cite{Almeida94}, making it particularly useful for analyzing chirp signals. 

\indent However, as explicitly pointed out by Stankovic and Sejdic in \cite{stankovic2019vertex}, the WD is not directly suitable for extension to GSP framework. Within GSP, the graph vertex--frequency energy distribution (GED) \cite{Ljub17,Stankovic20} has emerged as a more appropriate alternative, which serves as the graph counterpart of the classical Rihaczek distribution \cite{Rihaczek68,Scharf05}. The GED characterizes how the energy of a graph signal is distributed jointly over the vertex and frequency domains. Nevertheless, unlike the classical scenario, there is no corresponding rotational relationship between GED and GFRFT, primarily due to the non--Euclidean nature of graph domains and the generally non--idempotent property of the GFT matrix. As a result, the GED is not well-suited for analyzing graph chirp signals. To address this challenge, we introduce the graph fractional vertex--frequency energy distribution (GFED) by extending the GFT operator in GED to its fractional counterpart, the GFRFT. The GFED can capture the vertex--fractional-frequency representations, making it particularly suitable for nuanced analysis of graph chirp signals. Analogous to the fractional Rihaczek distribution in SP, the GFED can be regarded as a natural extension from traditional SP to GSP.

\indent When dealing with multi--component chirp signals, both the Rihaczek distribution and its fractional counterpart suffer from cross-term interference. These cross-terms not only obscure the true signal components but also significantly degrade the time--frequency resolution, hindering accurate interpretation. To address this issue, the Cohen's class time--frequency distribution (CD) \cite{Coh95} is developed as a comprehensive framework for constructing bi--linear kernel--based representations, with the Rihaczek distribution being one of its specific instances. Within this framework, extensive research has focused on designing adaptive kernel functions to effectively suppress cross-terms \cite{Zhao21,Rihacze68,Quan21,Zhao90,Zou22,Liang24}. The graph domain analogue of the CD is the general graph distribution (GGD). However, the GGD has similar limitations with GED that it struggles to effectively represent multi-component graph chirp signals. To overcome this limitation, we can replace the GFT operator within GGD with the GFRFT operator, obtaining the general fractional graph distribution (GFGD). This new distribution enhances the representation of graph chirp signals and improves cross-term suppression. Hence, the GFGD not only generalizes the fractional CD \cite{zhang2019cohen,Cui24,Cui2024generalized} to vertex--fractional-frequency domains in GSP but also constitutes a substantive extension of the GGD, tailored to address the cross-term interference of graph chirp signals.
} 

\begin{figure}
	\centering
	\includegraphics[scale=0.27]{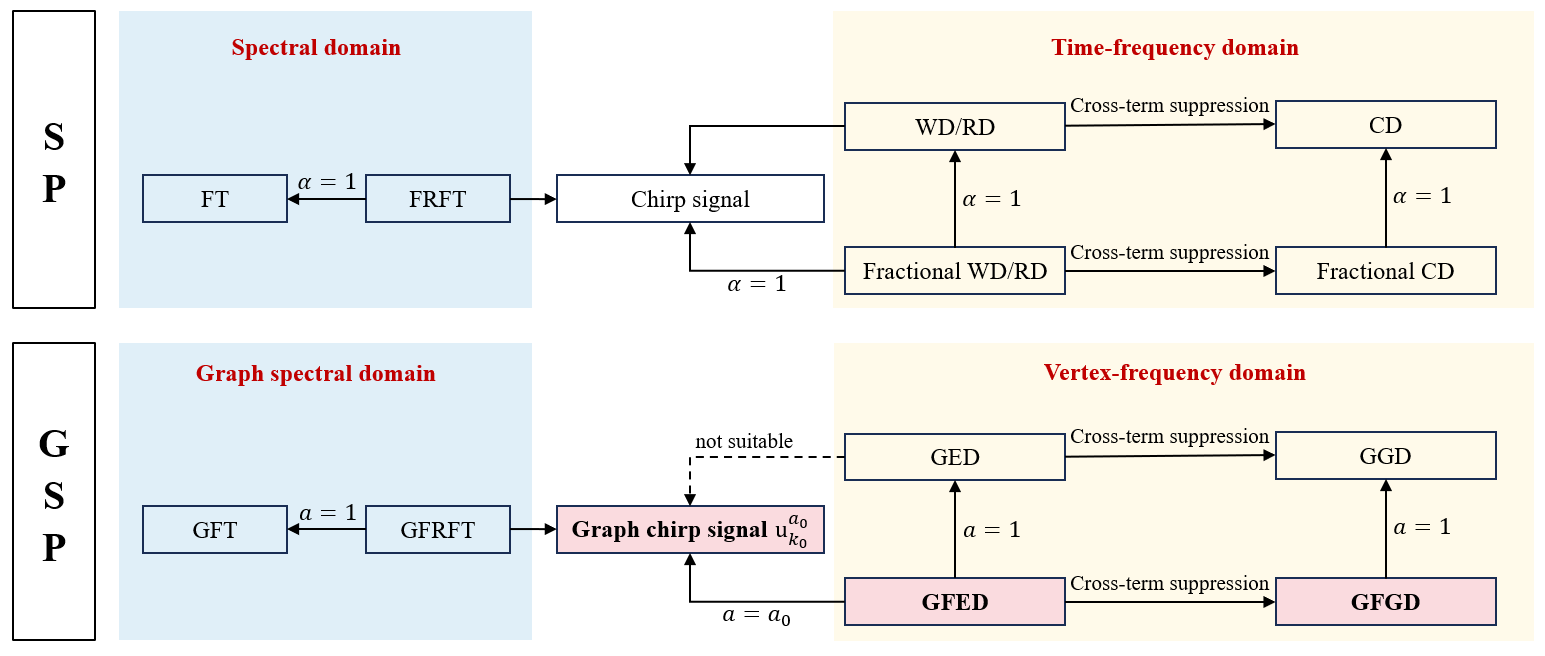}
	\caption{The framework and principal concepts of the paper, focusing on the relationship between SP and GSP.}
	\label{Fig_mind_map}
    \vspace{-0.5cm}
\end{figure}

\indent In summary, the analysis of graph chirp signals is crucial in GSP. The main objective of this paper is to define graph chirp signals and provide a tool for processing them by extending the GED and integrating it with the GFRFT to develop a fractionalized GED, thereby enhancing the theoretical framework for graph signal analysis. This approach offers an effective framework for analyzing and processing graph chirp signals with the potential to significantly impact the field of SP and graph theory. The main contributions of this paper are summarized as follows:
{
\begin{itemize}
	\item Define the \textbf{graph chirp signal} and derive its properties.
	
	\item Define the \textbf{GFED} and derive its properties.
	
	\item Define the \textbf{GFGD}, derive its properties, and provide some examples of reduced interference GFED.
	
	\item Develop a \textbf{GFED-based chirp detection method via filtering}, apply it to real-world denoising tasks, and achieve  superior performance to some state-of-the-art methods.
\end{itemize}
}
\indent  The remainder of this paper is organized as follows. Section \ref{sec2} introduces some preliminary concepts. Section \ref{sec3} defines graph chirp signals and discusses their relevant properties. Section \ref{sec4} presents the definition and properties of the GFED. Section \ref{sec5} introduces the GFGD and the reduced interference GFED. Section \ref{sec6} proposes a method for detecting graph chirp signals using filtering in the GFED domain. Section \ref{sec7} provides numerical experiments on the detection of graph chirp signals and the filtering of real-world data. Section~\ref{sec8} discusses modeling real--world signals with graph chirp signals and outlines possible extension. Finally, Section \ref{sec9} concludes the paper. Fig. \ref{Fig_mind_map} presents a mind map delineating the framework and principal concepts of this paper. All the technical proofs of our theoretical results are relegated to the Appendix parts.

\section{Preliminaries} \label{sec2}
\subsection{Graph Fourier transform}
\indent Let $\mathcal{G}=\left\{\mathcal{N},\mathcal{E},\mathbf{A}\right\}$ be a graph or network with a set of $N$ nodes $\mathcal{N}$, edges $\mathcal{E}$ such that $(m,n)\in\mathcal{E}$ and the graph adjacency $\mathbf{A}$. If there is no edge from node $n$ to node $m$, the element $\mathbf{A}_{mn}=0$, otherwise $\mathbf{A}_{mn} = 1$. The matrix $\mathbf{W}$ represents the strength of the connections between the nodes in the graph, where each element $W_{mn}$ denotes the weight of the edge between nodes $m$ and $n$. The diagonal elements $W_{nn}$ are zeros. The Laplacian is defined as $\mathbf{L=D-W}$, where $\mathbf{D}$ is the diagonal degree matrix with $D_{nn}=\sum_{m=1}^{N}W_{nm}$. For any undirected graph, the Laplacian is symmetric, i.e. $\mathbf{L}=\mathbf{L}^{\rm T}$. A graph signal $\mathbf{x}=[x(1),x(2),\cdots,x(N)]^{\mathrm{T}}$, $\mathbf{x} \in \mathbb{C}^{N}$, is defined as a mapping from the set of vertices $\mathcal{N}$ to $\mathbb{C}^{N}$, i.e. $\mathcal{N}_{n} \rightarrow x(n)$.

\indent In \cite{ortega22introduction}, the GFT is defined using the general graph shift operator (GSO) $\mathbf{Z}$ (e.g. adjacency $\mathbf{A}$, weighted adjacency $\mathbf{W}$, Laplacian $\mathbf{L}$, row normalized adjacency, summetric normalized Laplacian). In this paper, we considered the case where the underlying graph is undirected, so that the GSO $\mathbf{Z}$ is symmetric and diagonalizable. In particular, it can be decomposed as $\mathbf{Z} = \mathbf{U} \boldsymbol{\Lambda} \mathbf{U}^{-1}$. Here, $\mathbf{U}$ is the matrix whose columns are the eigenvectors $\mathbf{u}_{k}$, $k=1,2,\cdots,N$, and $\mathbf{\Lambda}$ is the diagonal matrix consisting of eigenvalues $\lambda_{1},\lambda_{2},\cdots,\lambda_{N}$ on the diagonal. It is obvious that $\mathbf{U}$ is unitary. For clarity, unless explicitly stated, the decomposition is based on the graph Laplacian in this paper. 

\indent The GFT of a signal $\mathbf{x}$ is defined as
\begin{align}
	\widehat{\mathbf{x}} = \mathbf{F}_{G}\mathbf{x} = \mathbf{U}^{-1}\mathbf{x},
\end{align}
where $\mathbf{F}_{G}=\mathbf{U}^{-1}$ is the GFT matrix and $\widehat{\mathbf{x}}=[\widehat{x}(1),\widehat{x}(2),\cdots,\widehat{x}(N)]$ denotes the GFT vector with $\widehat{x}(k)=\sum_{n=1}^{N}x(n)\overline{u_{k}(n)}$. Here, the superscript --- denotes the complex conjugate operator. 

\indent The inverse GFT is defined as
\begin{align}
	\mathbf{x} = \mathbf{F}_{G}^{-1} \widehat{\mathbf{x}} = \mathbf{U} \widehat{\mathbf{x}},
\end{align}
with $x(n) = \sum_{k=1}^{N}\widehat{x}(k)u_{k}(n)$.

\subsection{Graph fractional Fourier transform}
\indent By performing the Jordan decomposition on the GFT matrix $\mathbf{F}_{G}$, we express it as $\mathbf{F}_{G}=\mathbf{P} \mathbf{J}_{F} \mathbf{P}^{-1}$, where $\mathbf{P}$ is the matrix of generalized eigenvectors and $\mathbf{J}_{F}$ is the Jordan canonical form of $\mathbf{F}_{G}$. This decomposition allows us to obtain the GFRFT matrix $\mathbf{F}_{G}^{a}$ of order $a$ as follows:
\begin{align}
	\mathbf{F}_{G}^{a}= \mathbf{P} \mathbf{J}_{F}^{a} \mathbf{P}^{-1}.
\end{align}
\indent The $ath$ order GFRFT of any signal $\mathbf{x}$ is defined as \cite{Wang17}
\begin{align}
	\widehat{\mathbf{x}}_{a} = \mathbf{F}_{G}^{a} \mathbf{x},
\end{align}
where $\widehat{\mathbf{x}}_{a}=\left(\widehat{x}_{a}(1),\widehat{x}_{a}(2),\cdots,\widehat{x}_{a}(N)\right)$ is the GFRFT vector. It is obvious that the GFRFT matrices satisfy index additivity, that is $\mathbf{F}_{G}^{a}\mathbf{F}_{G}^{b}=\mathbf{F}_{G}^{a+b}$. Thus, the inverse GFRFT can be defined as
\begin{align}
	\mathbf{x} = (\mathbf{F}_{G}^{a})^{-1} \widehat{\mathbf{x}}_{a}.
\end{align}
\indent The GFRFT matrix $\mathbf{F}_{G}^{a}$ reduces to the unit matrix $\mathbf{I}_{N}$ for $a=0$, and the GFT matrix $\mathbf{F}_{G}$ for $a=1$.

\section{The graph chirp signal}  \label{sec3}
\indent LFM signal, also known as chirp signal, characterized by a frequency that varies with time, have long been utilized in classical SP for applications such as radar, communications, and system identification \cite{9121232,white2012performance,9525432}. A chirp signal typically exhibits a frequency that changes linearly with time. 

\indent In the continuous domain, a chirp signal is defined as \cite{Stank93,Djuric90}
\begin{align}  \label{LFM}
	f_{f_{0},f_{k}}(t) = {\rm e}^{{\rm i}(f_{0}t+\frac{f_{k}}{2}t^2)},
\end{align}
where $f_{0}$ and $f_{k}\neq 0$ denote the initial frequency and chirp rate, respectively. 

\indent We list several fundamental properties of chirp signals below.

\emph{Property 1 (FRFT--invariance)}  \label{property1}
Applying the FRFT to a chirp signal results in another chirp signal \cite{Shi17,Tao22}. When the chirp rate $f_k$ and the FRFT angle $\alpha$ satisfy the condition $f_k=-\mathrm{cot} \alpha$, the FRFT of the chirp signal becomes an impulse function at the frequency $w=\frac{f_0}{\mathrm{csc}^2 \alpha}$.

\emph{Property 2 (Constant amplitude)}  \label{property2}
Chirp signals have constant instantaneous amplitude. For all time $t\in \mathbb{R}$,
\begin{align}
	\left|f_{f_{0},f_{k}}(t)\right|^2=1 .
\end{align} 

\emph{Property 3 (Orthogonality)} \label{property3}
Chirp signals with different initial frequencies are orthogonal, , satisfying 
\begin{align}
	\langle f_{f_{0},f_{k}},f_{f_{1},f_{k}}\rangle=2\pi\delta_{f_{0},f_{1}},
\end{align}
where \( \langle \cdot, \cdot \rangle \) denotes the inner product and $\delta$ is the Kronecker delta function.

\emph{Property 4 (Time--frequency domain energy concentration)} \label{property4}
The fractional WD with order $\alpha$ \cite{Shi17,Tao22} of a chirp signal \( f_{f_0, f_k}(t) \) is given by
\begin{align}
	W^{\alpha}_f(t,w) 
	&= \int f\left(t+\frac{\tau}{2}\right) f^*\left(t-\frac{\tau}{2}\right) K_\alpha(\tau,w) d\tau \nonumber\\
	&= \sqrt{\rm i\, tan\alpha+1}\,\mathrm{e}^{\mathrm{i}\frac{1}{2}w^2\mathrm{cot}\alpha }\mathrm{e}^{-\mathrm{i}\frac{\left(f_0+f_kt-w\mathrm{csc}\alpha\right)}{2\mathrm{cot}\alpha}},
\end{align}
where $K_\alpha(\tau,w)=\sqrt{\frac{1-\mathrm{i\,cot}\alpha }{2\pi}}\mathrm{e}^{\mathrm{i}\frac{1}{2} \left(\tau^2+w^2\right)\mathrm{cot}\alpha-\mathrm{i} w\tau{\rm csc}\alpha}$ is the fractional kernel function. In particular, when ${\rm cot}\alpha = 0$, it reduces to the standard WD
\begin{align}
	W_f(t,w) = \sqrt{2\pi\left( 1-\mathrm{i\,cot}\alpha\right) }\,\delta\left(w - (f_0 + f_k t)\right),
\end{align}
which shows that the energy of a chirp signal lies along a line $w=f_0 + f_k t$ in the time--frequency plane.

\indent In SP, chirp signals are characterized by their instantaneous frequency being a linear function of time. However, this concept can not naturally extend to graph domains, where the vertex set is typically unordered and non--Euclidean, making it infeasible to define instantaneous frequency as a linear function of vertex index. Thus, a direct extension of the classical definition is not applicable. 

\indent Fortunately, as shown in Property 1, chirp signals also possess a critical inherent feature: they are the inverse FRFT of a delta function when the fractional order matches the chirp rate. Due to the reversibility of the FRFT, this means that chirp signals are uniquely characterized by the fact that their FRFT is a delta function, and vice versa. This feature is intrinsic to chirp signals.

\indent Motivated by this fundamental insight and the properties of classical chirp signals, we can extend the idea of chirp signals to the graph domain and define the graph chirp signal.

\indent \emph{Definition 1:} For a graph or network $\mathcal{G}$ with the GFT matrix $\mathbf{F}_{G}$, the graph chirp signal $\mathbf{u}_{k}^{a}$ ($a\neq0$) is defined as the inverse GFRFT with order $a$ of $\mathbf{e}_k$ as follows:
\begin{align}
	\mathbf{u}_{k}^{a} = \left(\mathbf{F}_{G}^{a}\right)^{-1}\mathbf{e}_k,
\end{align}   
where $a$ is the graph chirp rate, $k$ is the graph initial frequency and $\mathbf{e}_k$ denotes the standard basis vector with 1 at the $k$-th position and 0 elsewhere, i.e. $[\mathbf{e}_k]_i=\delta_{ik}$.   

\indent \emph{Theorem 1 (GFRFT--invariance):}  \label{thm1}
The graph chirp signal $\mathbf{u}_{k}^{a}$ remains a graph chirp under the GFRFT with order $b$, i.e.,
\begin{align}
	\widehat{\left(\mathbf{u}_{k}^{a}\right)}_b = \mathbf{u}_{k}^{a-b}.
\end{align}
\indent In particular, when $a=b$, we have
\begin{align}
	\widehat{\left(\mathbf{u}_{k}^{a}\right)}_a = \mathbf{e}_k.
\end{align}

\indent \emph{Proof:} See Appendix \ref{Appendix Theorems 1--4}.

\indent The GFRFT--invariance corresponds to the Property 1 of chirp signals. This stability under the GFRFT can also be interpreted as \textit{chirp rate additivity}, that is
\begin{align}
	\mathbf{u}_{k}^{a+b} = \mathbf{F}_{G}^{-b}\mathbf{u}_{k}^{a}.
\end{align}

\indent \emph{Theorem 2 (Constant--norm):}  \label{th2}
The graph chirp signals are unit-norm in the $\ell_2$-sense
\begin{align}
	\|\mathbf{u}_k^{a}\|_2 = 1.
\end{align}

\indent \emph{Proof:} See Appendix \ref{Appendix Theorems 1--4}.

\indent This implies that the energy of graph chirp signals is invariant with respect to the chirp rates and graph initial frequencies. Furthermore, this property complements Property 1 that since the GFRFT of order $b$ maps a graph chirp of rate $a$ to another graph chirp of rate $a-b$ and all such graph chirp signals are unit-norm, the total signal energy is preserved under successive GFRFT operations. Although GSP is inherently defined over discrete graph structures, the graph chirp rate is a continuous parameter. As a result, varying $a$ and $b$ continuously gives rise to a dense family of graph chirp signals and their transformations.  Therefore, the preservation of energy across these transformations can be interpreted as an \textit{approximate continuity} under GFRFT operations, despite the discrete nature of the underlying graph structure. The constant-norm of graph signals correspond to the Property 2 of chirp signals. 

\indent \emph{Theorem 3 (Orthogonality):}  \label{th3}
The graph chirp signals associated with different graph initial frequencies are orthogonal when the GFRFT matrix is unitary
\begin{align}
	\langle \mathbf{u}_k^{a}, \mathbf{u}_l^{a} \rangle = \delta_{kl}.
\end{align}

\indent \emph{Proof:} See Appendix \ref{Appendix Theorems 1--4}.

\indent It is obvious that when $k=l$, this identity reduces to the unit--norm condition established in Theorem \ref{th2}. Due to the orthogonality of the GFRFT basis, the set $\left\{ \mathbf{u}_k^{a},k=1,2,\cdots,N \right\}$ forms an orthonormal basis, enabling graph signals to be efficiently decomposed, projected, and analyzed in the graph spectral domain. This result parallels the Property 3 of chirp signals, which asserts orthogonality between chirp signals of different initial frequencies.

\begin{figure}  
	\centering
	\includegraphics[scale=0.45]{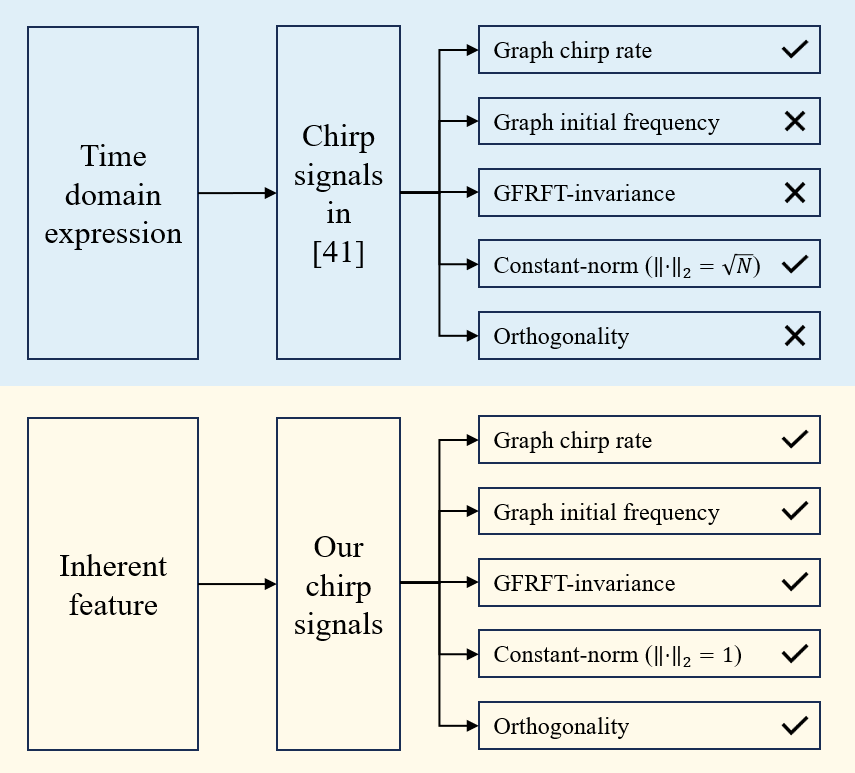}
	\caption{The comparison of two kind of graph chirp signals.}
	\label{Fig_chirp_signals_compare}
	\vspace{-0.5cm}
\end{figure}

\indent \emph{Comparisons with the graph chirp signals defined in \cite{chen2024graph}:} A related concept of graph chirp signal was previously introduced by Chen et al. in their work \cite{chen2024graph}. Rather than generalizing the definition in time domain directly --- which is infeasible due to the lack of a natural vertex ordering --- they define the graph chirp signals from the perspective of graph spectral domain. Their method emulates the spectral form in the GFT domain of classical chirp signals, with the objective of approximating the time domain expression described in \eqref{LFM}. In contrast, our approach originates from the inherent properties of chirp signals rather than their explicit formula. We define graph chirp signals as the inverse GFRFT of a spectral delta function,  ensuring that they inherit the core characteristics of chirp signals. Notably, the graph chirp signal defined by Chen et al. involves only the graph chirp rate, without incorporating a notion of graph initial frequency. In contrast, our definition allows for both the graph chirp rate and graph initial frequency, mirroring the dual-parameter structure of classical chirp signals. Furthermore, their graph chirp signal does not satisfy the GFRFT-invariance, and due to the absence of graph initial frequency, it also fails to exhibit orthogonality. The comparison of two kinds of graph chirp signals is illustrated in Fig. \ref{Fig_chirp_signals_compare}.

\indent  The adjacency matrix of a cycle graph is given by
\begin{align}
	\mathbf{A}=\left[\begin{array}{llll}
		  & & & 1 \\
		1 & & &   \\
		& \ddots & & \\
		& & 1 &
	\end{array}\right].
\end{align}
On a cycle graph, the GFRFT matrix, $\mathbf{F}_{G}^{a}$, is equal to the  discrete fractional Fourier transform (DFRFT) matrix \cite{Wang17}. Consequently, the graph chirp signals on a cycle graph are defined as the columns of the DFRFT matrix with a fractional order $-a$.

\begin{figure}[!t]
	\centering
	\includegraphics[width=0.47\textwidth]{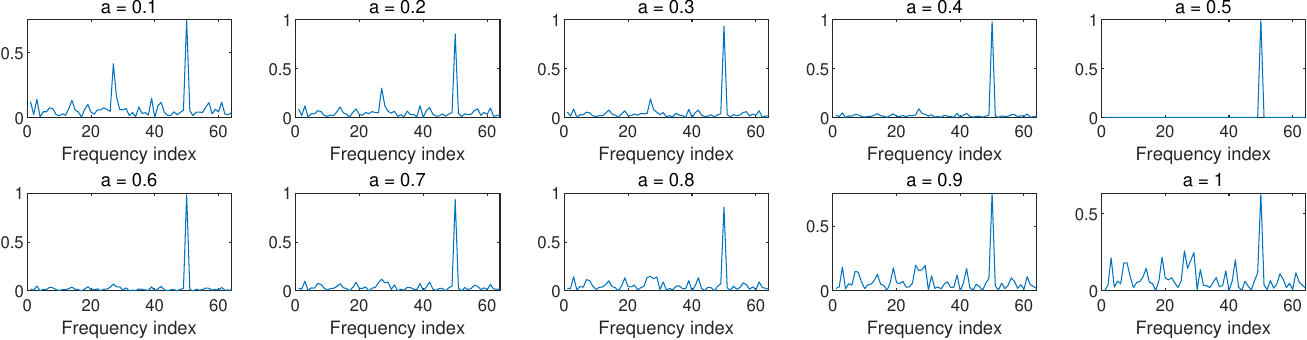}
	\caption{GFRFTs of the chirp signal $\mathbf{u}_{50}^{0.5}$ on David sensor at different fractional orders.}
	\label{Fig03}
	\vspace{-0.5cm}
\end{figure}

\indent The GFRFT of a graph chirp signal $\mathbf{u}_{k}^{a}$ produces a corresponding impulse in the graph fractional spectral domain, analogous to the chirp signal's behavior in classical SP. For simplicity, let $\mathbf{U}_{a}=(\mathbf{u}_{1}^{a},\mathbf{u}_{2}^{a},\cdots,\mathbf{u}_{N}^{a})$. For example, Fig. \ref{Fig03} illustrates the GFRFT of the chirp signal $\mathbf{u}_{50}^{0.5}$ on the David sensor, with the fractional order $a$ varying from $0.1$ to $1$. It can be seen that the GFRFT for this fractional order generates an impulse at the chirp rate $a=0.5$, indicating that the signal exhibits a well-defined frequency concentration at this order. At other fractional orders, the spectrum is more mixed. The GFRFT effectively handles graph chirp signals by providing a flexible method to adjust the order. This aligns with the classical treatment of chirp signals, where the frequency evolution over time is captured precisely.

\section{Fractional vertex--frequency energy distribution} \label{sec4}
\subsection{Vertex--frequency energy distribution}
\indent In classical SP, the WD \cite{Wig32,Sad21} is a well-known tool that provides a joint representation of a signal, offering valuable insight into its time--frequency characteristics. It is particularly effective for the analysis and processing of chirp signals, such as filtering and detection. In GSP, the concept of energy distribution holds an analogous importance to the role of WD in classical SP. 

\indent In \cite{Ljub17}, the GED is proposed, which corresponds to the Rihaczek distribution in time--frequency analysis and is defined as follows: 
\begin{align}
	E_{\mathbf{x}}(n,k) = x(n)\thinspace\overline{\widehat{x}(k)}\thinspace \overline{u_{k}(n)}.
\end{align}
\indent It satisfies the marginal properties:
\begin{align}
	\sum_{n=1}^{N}E_{\mathbf{x}}(n,k) = \left|\widehat{x}(k)\right|^2,
\end{align}
\begin{align}
	\sum_{k=1}^{N}E_{\mathbf{x}}(n,k) = |x(n)|^2.
\end{align}

\subsection{Fractional vertex--frequency energy distribution}
\indent The energy of a signal $\mathbf{x}$ in classical SP is
\begin{align}
	E_{\mathbf{x}}=\sum_{n=1}^{N}|x(n)|^2.
\end{align} 
\indent To preserve the energy, the above equation can be rewritten through GFRFT as
\begin{align}
	E_{\mathbf{x}}=\sum_{n=1}^{N}\sum_{k=1}^{N}x(n)\overline{\widehat{x}_{a}(k)}\thinspace \overline{u_{k}^{a}(n)}.
\end{align}

\indent Thus, we can define the GFED to further enhance the flexibility of signal analysis on graphs. In particular, since the GFT operator used in the GED is not well--suited for handling graph chirp signals, we propose replacing it with the GFRFT operator, which is more appropriate for analyzing such signals. This substitution allows the GFED to better capture the joint vertex–fractional-frequency characteristics of graph chirps, thereby further enhancing the flexibility and effectiveness of signal analysis on graphs.

\indent \emph{Definition 2:} For the signal $\mathbf{x}$ defined on the graph $\mathcal{G}$, the GFED $E_{\mathbf{x}}^{a}$ of order $a$ is defined as
\begin{align}
	E_{\mathbf{x}}^{a}(n,k) =& x(n)\thinspace\overline{\widehat{x}_{a}(k)}\thinspace \overline{u_{k}^{a}(n)} \nonumber\\
	=& \sum_{m=1}^{N} x(n)\thinspace\overline{x(m)}\thinspace \overline{u_{k}^{a}(n)}\thinspace u_{k}^{a}(m) .
\end{align}

\indent When $a=1$, the GFED reduces to the GED. Obviously, the GFED satisfies the marginal properties as well.

\indent \textbf{Vertex marginal property:}
\begin{align}
	\sum_{k=1}^{N}E_{\mathbf{x}}^{a}(n,k) = |x(n)|^2.
\end{align}
\indent \textbf{Frequency marginal property:}
\begin{align}
	\sum_{n=1}^{N}E_{\mathbf{x}}^{a}(n,k) = \left|\widehat{x}_{a}(k)\right|^2.
\end{align}

\indent \emph{Theorem 4 (Vertex--fractional-frequency domain energy concentration):}  \label{th4}
Let $\mathbf{u}_{k_0}^{a_0}$ be a graph chirp signal with graph chirp rate $a_0$ and graph initial frequency $k_0$. Then its GFED with order $a$ satisfies
\begin{align}  \label{Fractional energy concentration1}
	E_{\mathbf{u}_{k_0}^{a_0}}^a(n,k) &= u_{k_0}^{a_0}(n)\,\overline{\widehat{\left( u_{k_0}^{a_0}\right)}_a(k)}\,\overline{u_{k}^{a}(n)} \nonumber\\
	&=u_{k_0}^{a_0}(n)\,\overline{ u_{k_0}^{a_0-a}(k)}\,\overline{u_{k}^{a}(n)} .
\end{align}
\indent In particular, when the graph chirp rate is equal to the fractional order, i,e, $a_0=a$, we have
\begin{align}  \label{Fractional energy concentration2}
	E_{\mathbf{u}_{k_0}^{a}}^a(n,k) = \begin{cases}
		\left| \mathbf{u}_{k_0}^{a}(n) \right|^{2}    & k=k_0 \\
		0  & k\neq k_0
	\end{cases}.
\end{align}

\indent \emph{Proof:} See Appendix \ref{Appendix Theorems 1--4}.

\indent Notably, while GED fail to exhibit energy concentration for graph chirp signals, the GFED achieves perfect localization when the chirp rate $a_0$ matches the GFED order $a$, all energy is concentrated at the fractional frequency $k_0$. This result corresponds to the Property 4 of chirp signals. However, as explicitly pointed out by Stankovic and Sejdic in \cite{stankovic2019vertex}, the WD is not suitable for the graph framework extension. Therefore, in the graph setting, we consider the GFED as a functional analogue of the fractional WD/RD.

\indent Although in classical SP the FRFT is widely interpreted as a rotation operator in the time–-frequency plane, effectively rotating the WD. However, this geometric interpretation can not extend to GSP, which operates in a non-Euclidean domain where concepts such as rotation are not naturally defined. The reason the FRFT can be interpreted as a rotation operator stems from the fact that the FT matrix $\mathbf{F}$ is a power--idempotent operator, satisfying $\mathbf{F}^4 = \mathbf{I}_{N}$. In contrast, the GFT matrix is generally not power-idempotent, due to the irregularity of the underlying graph structure. If there exists an integer $a$ such that the GFT matrix satisfies $(\mathbf{F}_G)^a = \mathbf{I}_N$, the integer $a$ can be interpreted as a form of generalized periodicity. However, such cases are rare. Therefore, the GFRFT does not produce a rotational effect analogous to that in the Euclidean setting, and the notion of rotation in the vertex–-frequency domain is not applicable in the framework of GSP.

\begin{figure}[!t]
	\centering
	\subfloat[]{\includegraphics[width=1.8in]{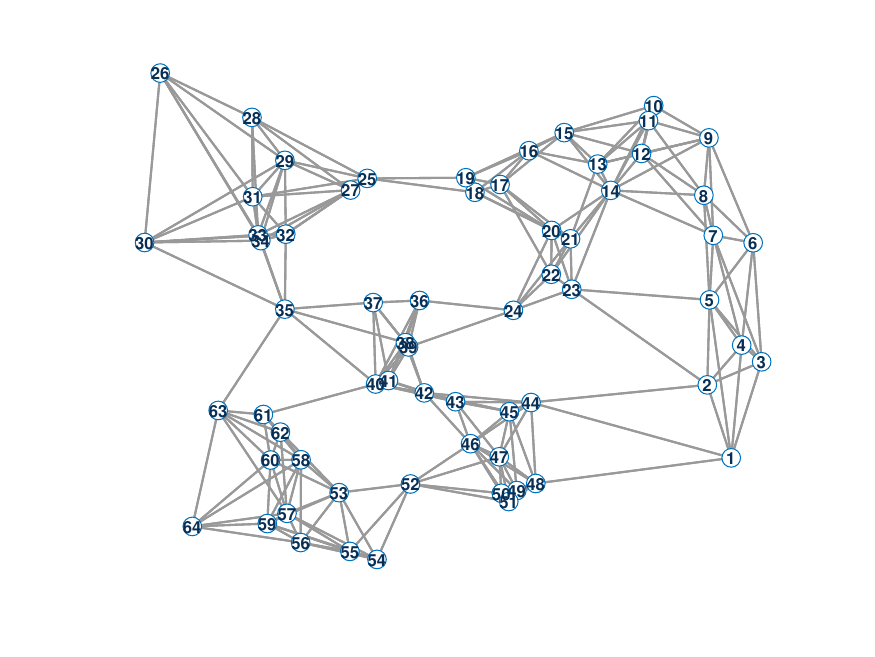}%
			\label{4a}}
			
	\begin{minipage}{\linewidth}
		\subfloat[]{\includegraphics[width=1.7in]{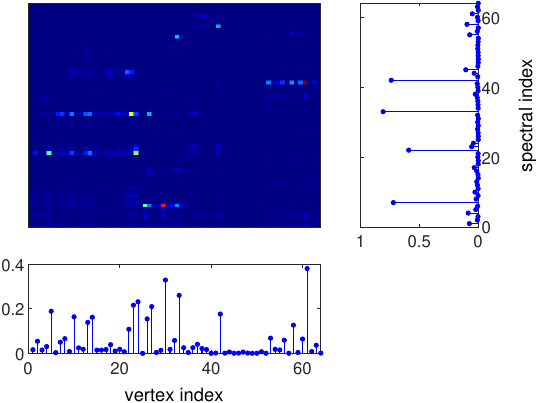}%
			\label{4b}} 
	\hspace{\fill}
		\subfloat[]{\includegraphics[width=1.7in]{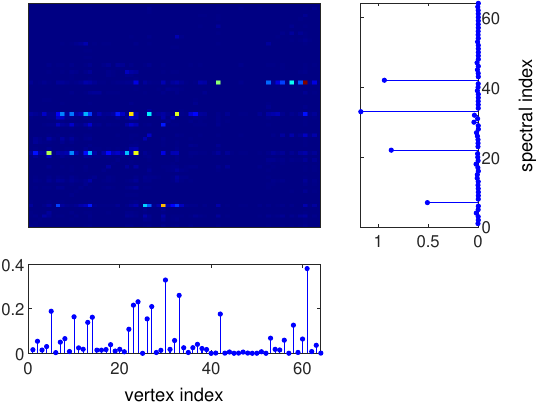}%
			\label{4c}}
	\end{minipage}
	\vspace{-0.1cm}
	\caption{Sensor network graph structure and GFEDs. (a) Graph structure. (b) The GED of $\mathbf{x}_{1}$. (c) The GFED of $\mathbf{x}_{1}$ ($a=0.7$). }
	\label{Fig4}
	\vspace{-0.4cm}
\end{figure}

\indent \emph{Example 1:} Consider a sensor network graph presented in Fig. \ref{Fig4}\subref{4a}. The graph chirp frequency $a=0.7$ is selected for the graph chirp signals, with initial frequencies chosen as follows. The graph signal $\mathbf{x}_{1}$, is defined on vertices 1 to 24 using $\mathbf{u}_{22}^{a}$, on vertices 25 to 34 using $\mathbf{u}_{7}^{a}$, and on vertices 35 to 64 using $\mathbf{u}_{42}^{a}$, along with the additional $\mathbf{u}_{33}^{a}$. The GED and GFED, along with the marginal properties are illustrated in Fig. \ref{Fig4}. 

\begin{figure}[!t]
	\centering
	\subfloat[]{\includegraphics[width=1.8in]{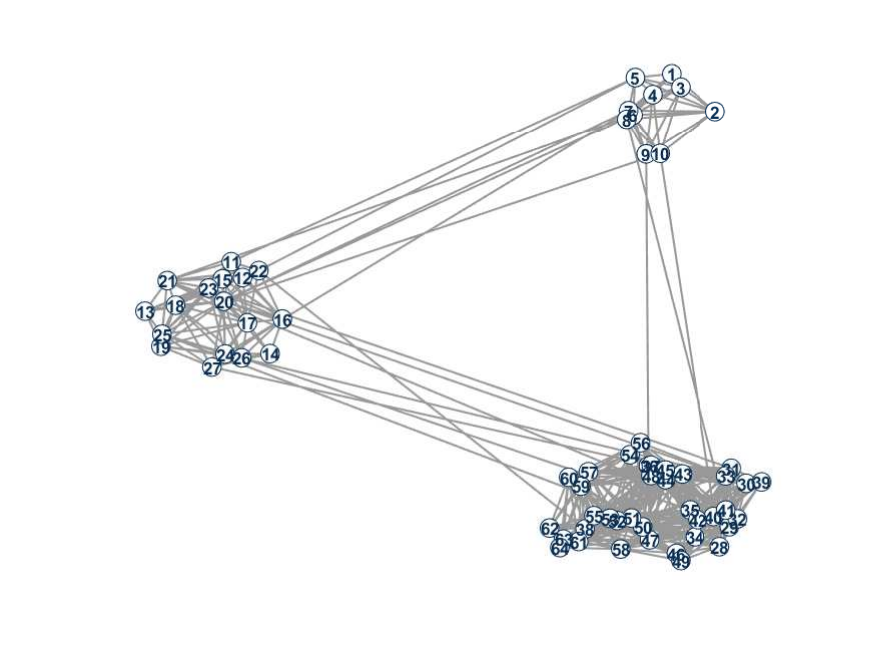}%
		\label{5a}}
	
	\begin{minipage}{\linewidth}
	\subfloat[]{\includegraphics[width=1.7in]{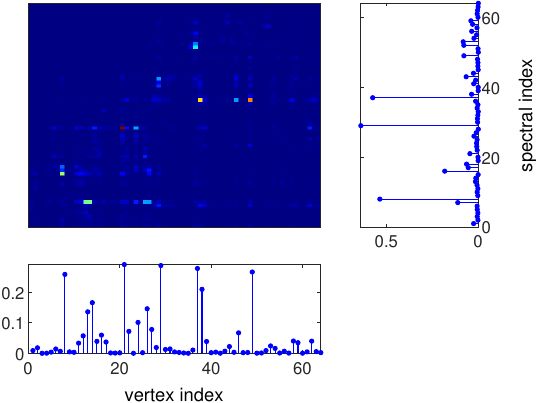}%
		\label{5b}}
	\hspace{\fill}
	\subfloat[]{\includegraphics[width=1.7in]{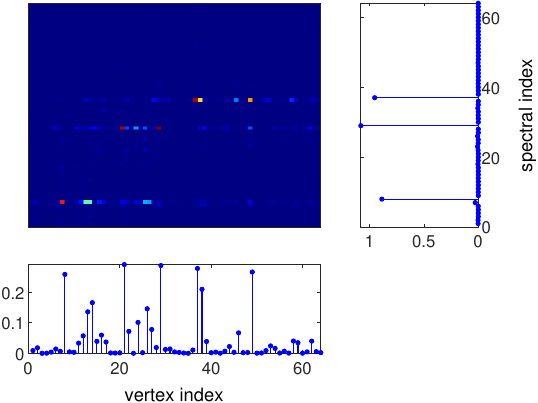}%
		\label{5c}}
	\end{minipage}
	\vspace{-0.1cm}
	\caption{Community network graph structure and GFEDs. (a) Graph structure. (b) The GED of $\mathbf{x}_{2}$. (c) The GFED of $\mathbf{x}_{2}$ ($a=0.6$).}
	\label{Fig5}
	\vspace{-0.6cm}
\end{figure}

\indent \emph{Example 2:} Consider a community network graph presented in Fig. \ref{Fig5}\subref{5a}. The graph chirp frequency $a=0.6$ is selected for the graph chirp signals, with initial frequencies chosen as follows. The graph signal $\mathbf{x}_{2}$, is defined on vertices 1 to 27 using the graph chirp signal $\mathbf{u}_{8}^{a}$, and on vertices 28 to 64 using $\mathbf{u}_{37}^{a}$, along with an additional $\mathbf{u}_{29}^{a}$. The GED and GFED, along with the marginal properties are illustrated in Fig. \ref{Fig5}. 

\indent As observed in Figs. \ref{Fig4} and \ref{Fig5}, the GFED representations offer clearer and more concentrated characterizations of graph chirp signals compared to the GED. Taking Example 2 as illustrated in Fig. \ref{Fig5}, the difference between the two representations is evident. In Fig. \ref{Fig5}\subref{5b}, the GED of $\mathbf{x}_2$ displays several spread-out spectral components, with activations not only around the true initial frequencies (indices 8, 29, and 37), but also in unrelated regions such as around index 20,  and between indices 40--60. These additional activations indicate spectral leakage and reduce interpretability. By contrast, as shown in Fig. \ref{Fig5}\subref{5c}, the GFED representation of $\mathbf{x}_2$ (with $a=0.6$) demonstrates a much more concentrated spectral profile, with energy sharply localized around the expected frequencies. This indicates that GFED provides a more accurate and compact vertex--fractional-frequency representation of graph chirp signals.

\indent To move beyond purely visual comparison, we introduce Shannon entropy to quantify the concentration of energy. Given a normalized vertex--frequency distribution $\mathbf{D}_\mathbf{x}=\left[D(n,k)\right]_{N\times N}$ such that $\sum_n\sum_k\left|D(n,k)\right|^2=1$, the Shannon entropy is defined as \cite{stankovic2001measure}
\begin{align}
	SE\left(\mathbf{D_x}\right)=-\sum_{n}\sum_{k}\left|D(n,k)\right|\mathrm{log}_2 \left|D(n,k)\right|.
\end{align}
A lower entropy value implies that the energy is more concentrated, while a higher entropy indicates a more dispersed energy distribution. Table \ref{Shannon-GFED} shows the entropy values for both GEDs and GFEDs of $\mathbf{x}_1$ and $\mathbf{x}_2$. As presented in the Table, GFED consistently yields lower entropy values.

\indent It has been demonstrated in Examples 1 and 2 that the GEDs fail to adequately capture the characteristics of graph chirp signals. In contrast, the GFEDs provide superior performance, effectively revealing vertex--fractional-frequency information, accurately identifying initial frequency components, and achieving greater energy concentration. 

\begin{table}[!t]
	\caption{Shannon entropy values of GEDs and GFEDs. }\label{Shannon-GFED}
	\centering
	\begin{tabular}{c|c|c}
		\hline
		Signal & Distribution & Shannon entropy\\ \hline
		\multirow{2}{*}{$\mathbf{x}_1$} & GED & $55.8646$ \\ 
		& GFED ($a=0.7$) & $\textbf{54.5265}$\\ \hline
		\multirow{2}{*}{$\mathbf{x}_2$} & GED & $64.4671$ \\
		& GFED ($a=0.6$) & $\textbf{30.6718}$\\
		\hline
	\end{tabular}
	\vspace{-0.3cm}
\end{table}

\section{General Fractional Graph Distribution}  \label{sec5}
\indent The GGD form is \cite{Sta18}
\begin{align}
	G_{\mathbf{x}}(n,k)=\sum_{p=1}^{N}\sum_{q=1}^{N}\widehat{x}(p) \overline{\widehat{x}(q)} u_{p}(n)\overline{u_{q}(n)}\phi(p,k,q),
\end{align}
where $\phi$ is the kernel function.

\indent When $\phi(p,k,q)=\delta(q-k)$, the GGD reduces to the GED. Additionally, if the kernel satisfies the condition $\sum_{k=1}^{N}\phi(p,k,q)=1$, the unbiased condition holds, i.e. $\sum_{n=1}^{N}\sum_{k=1}^{N}G_{\mathbf{x}}(n,k)=E_{\mathbf{x}}$.

\indent However, the GGD is also not well--suited for processing graph chirp signals. To address this limitation, we introduce the GFGD.

\indent \emph{Definition 3:} For the signal $\mathbf{x}$ defined on the graph $\mathcal{G}$, the GFGD $G_{\mathbf{x}}^{a}$ of order $a$ is defined as
\begin{align}  \label{GFGD1}
	G_{\mathbf{x}}^{a}(n,k)=\sum_{p=1}^{N}\sum_{q=1}^{N}\widehat{x}_{a}(p) \overline{\widehat{x}_{a}(q)} u^{a}_{p}(n)\overline{u^{a}_{q}(n)}\phi(p,k,q).
\end{align}

\indent It is evident that when $a=1$, the GFGD reduces to the GGD. Moreover, when $\phi(p,k,q)=\delta(q-k)$, the GFGD reduces to the GFED. Additionally, the unbiased energy condition  $\sum_{n=1}^{N}\sum_{k=1}^{N}G_{\mathbf{x}}^{a}(n,k)=E_{\mathbf{x}}$ holds if $\sum_{k=1}^{N}\phi(p,k,q)=1$. 

\indent The GFGD satisfies the vertex and frequency marginal properties as follows:

\indent \textbf{Vertex marginal property:}\\
\indent If $\sum_{k=1}^{N}\phi(p,k,q)=1$, the GFGD satisfies the vertex marginal property
\begin{align}
	\sum_{k=1}^{N}G_{\mathbf{x}}^{a}(n,k)&=\sum_{p=1}^{N}\sum_{q=1}^{N}\widehat{x}_{a}(p) \overline{\widehat{x}_{a}(q)} u^{a}_{p}(n)\overline{u^{a}_{q}(n)} \nonumber\\
	&=\left|x(n) \right|^2.
\end{align}
\indent Also, the vertex moment property
\begin{align}
	\sum_{n=1}^{N}\sum_{k=1}^{N}n^{m}G_{\mathbf{x}}^{a}(n,k)=\sum_{n=1}^{N}n^{m}\left|x(n) \right|^2
\end{align} 
holds for the same condition.

\indent \textbf{Frequency marginal property:} \\
\indent If $\phi(p,k,p)=\delta(p-k)$, the GFGD satisfies the frequency marginal property
\begin{align}
	\sum_{n=1}^{N}G_{\mathbf{x}}^{a}(n,k)
	=\sum_{p=1}^{N} \left|\widehat{x}_{a}(p) \right|^2 \phi(p,k,p)
	= \left|\widehat{x}_{a}(k) \right|^2.
\end{align}
\indent Also, the frequency moment property
\begin{align}
	\sum_{n=1}^{N}\sum_{k=1}^{N}n^{m}G_{\mathbf{x}}^{a}(n,k)=\sum_{k=1}^{N}k^{m}\left|\widehat{x}_{a}(k) \right|^2
\end{align} 
holds for the same condition.

\indent Additionally, the GFGD can be rewritten as a dual form of \eqref{GFGD1}, expressed in the vertex--vertex domain as
\begin{align}  
	G_{\mathbf{x}}^{a}(n,k)=\sum_{m=1}^{N}\sum_{t=1}^{N}x(m) \overline{x(t)} u^{a}_{k}(m)\overline{u^{a}_{k}(t)}\varphi(m,n,t),
\end{align}
where $\varphi(m,n,t)$ is the kernel function in this domain, which has the same mathematical form with the frequency--frequency domain kernel in essence. Furthermore, if $\varphi(m,n,t)=\delta(m-n)$, the vertex marginal property is satisfied. Similarly, if $\sum_{n=1}^{N}\varphi(m,n,t)=1$, the frequency marginal property is satisfied.

\indent The form of Choi-Williams kernel in GSP is \cite{Sta18}
\begin{align}
	\phi(p,k,q)=\frac{1}{s(p,q)}{\rm e}^{-\gamma\frac{|\lambda_{k}-\lambda_{q}|}{|\lambda_{p}-\lambda_{q}|}},
\end{align}
where $s(p,q)=\sum_{k=1}^{N}{\rm e}^{-\gamma\frac{|\lambda_{k}-\lambda_{q}|}{|\lambda_{p}-\lambda_{q}|}}$ for $p\neq q$ and $s(p,q)=\delta(p-q)$ for $p=q$. The reduced interference GFED using the Choi-Williams kernel satisfies the marginal properties. 

\begin{figure}[!t]
	\centering
	\subfloat[]{\includegraphics[width=1.7in]{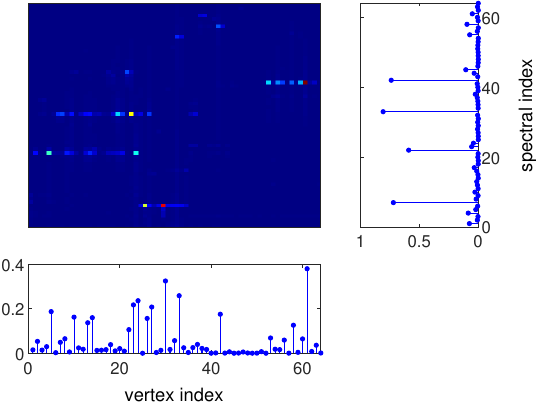}%
		\label{6a}}
	\hspace{\fill}
	\subfloat[]{\includegraphics[width=1.7in]{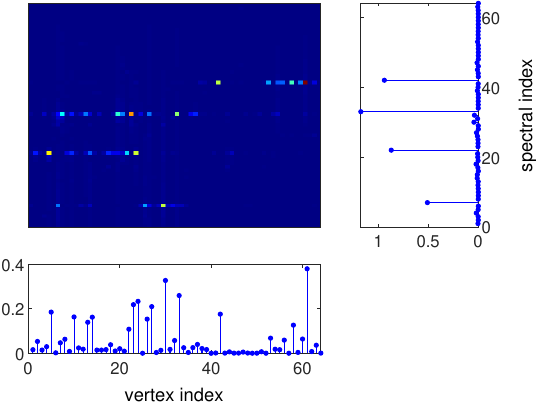}%
		\label{6b}}
	\
	\subfloat[]{\includegraphics[width=1.7in]{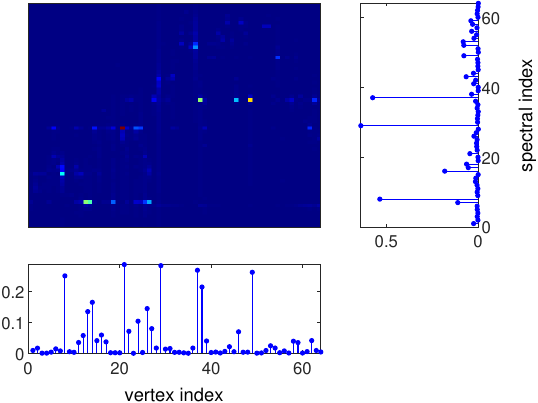}%
		\label{6c}} 
	\hspace{\fill}
	\subfloat[]{\includegraphics[width=1.7in]{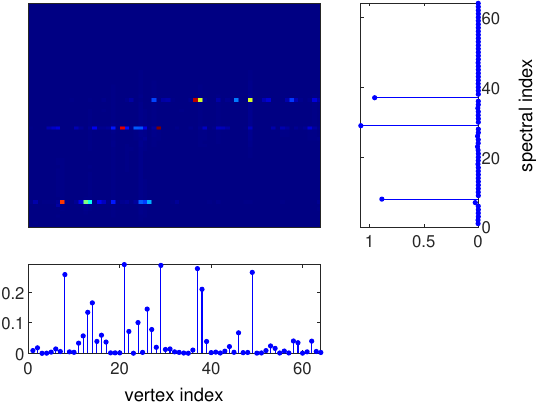}%
		\label{6d}} 
	\vspace{-0.1cm}
	\caption{The GGD-CWs and GFGD-CWs of the two signals. (a) The GGD-CW of $\mathbf{x}_{1}$. (b) The GFGD-CW of $\mathbf{x}_{1}$. (c) The GGD-CW of $\mathbf{x}_{2}$. (d) The GFGD-CW of $\mathbf{x}_{2}$.}
	\label{Fig6}
	\vspace{-0.5cm}
\end{figure}

\indent For the signals used in Examples 1--2, Fig. \ref{Fig6} plots the GGD with Choi-Williams kernel (GGD-CW) and GFGD with Choi-Williams kernel (GFGD-CW). As observed from Figs. \ref{Fig6}\subref{6a} and \ref{Fig6}\subref{6c}, the GGD-CW helps reduce interference in the GED. However, it still fails to accurately capture the true frequency components, as spectral artifacts remain and interfere with the interpretation. In contrast, the GFGD-CW achieves superior suppression of interference in the GFEDs and provides much clearer representations of the vertex--fractional-frequency features, offering a more accurate depiction of the signals. In addition, Table \ref{Shannon-GFGD-CW} shows the corresponding Shannon entropy values. The GFGD-CW consistently yields lower entropy values than the GGD-CW, indicating higher energy concentration and a more compact representation. These quantitative results are consistent with the visual analysis and confirm the effectiveness of GFGD-CW in reducing interference in the vertex--fractional-frequency domains.

\begin{table}[!t]
	\caption{Shannon entropy values of GGD-CWs and GFGD-CWs. }\label{Shannon-GFGD-CW}
	\centering
	\begin{tabular}{c|c|c}
		\hline
		Signal & Distribution & Shannon entropy\\ \hline
		\multirow{2}{*}{$\mathbf{x}_1$} & GGD-CW & $55.5006$ \\ 
		& GFGD-CW ($a=0.7$) & $\textbf{48.2836}$\\ \hline
		\multirow{2}{*}{$\mathbf{x}_2$} & GGD-CW & $57.2824$ \\
		& GFGD-CW ($a=0.6$) & $\textbf{31.2431}$\\
		\hline
	\end{tabular}
	\vspace{-0.3cm}
\end{table}

\section{Detection of Graph Signals through Fractional vertex--frequency energy Distribution domain Filtering}  \label{sec6}
\indent Filtering and detecting play crucial roles in GSP, as they help in extracting relevant information while reducing unwanted noise. Typically, filtering is performed either in the vertex domain or in the frequency domain. In this section, we focus on signal detection through filtering techniques in the vertex--fractional-frequency domain, a hybrid approach that combines the strengths of both the vertex and frequency domains.

\indent For the pure and noisy signals $\mathbf{x}$ and $\mathbf{y}$, we use the signal observation model $\mathbf{y} = \mathbf{x}+\mathbf{w}$, where $\mathbf{w}$ represents the additive Gaussian noise. According to Wiener filter principle, a natural criterion to characterize the estimation accuracy is the MSE criterion
\begin{align}  \label{MSE}
	\mathrm{MSE} = \mathbb{E}\left \{\left \| E^{a}_{\mathbf{x}}-E_{\widetilde{\mathbf{x}}}^{a} \right \|_{F}^{2}\right \} .
\end{align}
where $\mathbb{E}$ denotes the mathematical expectation operator and $E_{\widetilde{\mathbf{x}}}^{a}$ is the estimated GFED. We can then formulate the problem as 
\begin{align}
	\sigma^{2}_{\mathrm{MSE}} = \min_{\mathbf{H}} \mathrm{MSE}.
\end{align}

\indent Since the GFT of vertex--frequency distributions has not been previously defined, firstly,  we need to define the GFT of GFED. For any signal $\mathbf{x}$, the GFED $E_{\widetilde{\mathbf{x}}}^{a}$ can be considered as a consist of $N$ signals  $E_{\widetilde{\mathbf{x}}}^{a}(n,k)$, where $k=1,2,\cdots,N$. Thus, we define the GFT of GFED as
\begin{align}
	\widehat{E_{\mathbf{x}}^{a}}=\mathbf{U}^\mathrm{H}E_{\mathbf{x}}^{a},
\end{align}
and its inverse as
\begin{align}   \label{inverse GFED  GFT}
	E_{\mathbf{x}}^{a} = \mathbf{U} \widehat{E_{\mathbf{x}}^{a}} .
\end{align}

\indent Similar to the convolution theorem in classical Fourier analysis, in GSP, we can utilize the convolution operator $\star$ to model the filtering process. Specifically, the estimated GFED $E_{\widetilde{\mathbf{x}}}^{a}$ is obtained through the convolution of the GFED of the noisy signal $\mathbf{y}$ and the filter $\mathbf{H}$ in vertex--fractional-frequency domain as
\begin{align}  \label{GFED estimate}
	E_{\widetilde{\mathbf{x}}}^{a}=E_{\mathbf{y}}^{a}\star\mathbf{H}
	= \mathbf{U}(\widehat{E_{\mathbf{y}}^{a}}\circ\widehat{\mathbf{H}}) ,
\end{align}
where $\circ$ denotes Hadamard product.

\indent \emph{Theorem 5:}  \label{Th5}
For the MSE minimization problem in \eqref{MSE}, if the original signal $\mathbf{x} \in \mathbb{C}^N$ is deterministic, the noise $\mathbf{w}$ is zero-mean, circular complex Gaussian with variance $\sigma^2$, and $\mathbf{x}$ and $\mathbf{w}$ are statistically independent, then the filter transfer function matrix in the frequency--fractional-frequency domain that minimizes the MSE as
\begin{align}  \label{filter transfer matrix}
	\widehat{\mathbf{H}} = \left(\mathbb{E}\left\{\widehat{E^{a}_{\mathbf{y}}}\circ\overline{\widehat{E^{a}_{\mathbf{y}}}}\right\}\right)^{-\circ}\circ\left(\widehat{E_{\mathbf{x}}^{a}}\circ\mathbb{E}\left\{\overline{\widehat{E^{a}_{\mathbf{y}}}}\right\}\right).  
\end{align}

\indent \emph{Proof:} See Appendix \ref{Appendix filter transfer matrix}.

\indent \emph{Theorem 6:}  \label{Th6}
	The optimal filter in the vertex--fractional-frequency domain for the MSE minimization problem in \eqref{MSE} can be expressed as \eqref{optimal filter}.
\begin{figure*}[htbp]
	\begin{align}    \label{optimal filter}
		H(n,k) = \sum_{l}\frac{
			\left[{\left|\widehat{E^{a}_{\mathbf{x}}}(l,k)\right|^2 + \sigma^2 \widehat{E^{a}_{\mathbf{x}}}(l,k) \sum_{i=1}^{N} \left|U_{a}(i,k)\right|^2 \overline{U(i,l)} }\right] U(n,l)
		}{
			\begin{aligned}[t]
				&\Bigg[ \left|\widehat{E^{a}_{\mathbf{x}}}(l,k)\right|^2 
				+ \sigma^2 \sum_{i}\sum_{j} x(i) \overline{\widehat{x}_{a}(k)} \overline{U_{a}(i,k)} \left|U_{a}(j,k)\right|^2 \overline{U(i,l)} U(j,l) 
				+ \sigma^2 \left|\sum_{i} x(i) \overline{U_{a}(i,k)} \overline{U(i,l)} \right|^2 \\
				&+ \sigma^2 \sum_{i} \left|\widehat{x}_{a}(k)\right|^2 \left|U_{a}(i,k)\right|^2 \left|U(i,l)\right|^2 
				+ \sigma^2 \sum_{i}\sum_{j} \overline{x(j)} \widehat{x}_{a}(k) \left|U_{a}(i,k)\right|^2 U_{a}(j,k) \overline{U(i,l)} U(j,l) \\
				&+ \sigma^4 \left|\sum_{i} \left|U_{a}(i,k)\right|^2 U(i,l)\right|^2 
				+ \sigma^4 \sum_{i} \left|U_{a}(i,k)\right|^2 \left|U(i,l)\right|^2 
				+ 2 \sigma^4 \sum_{i} \left|U_{a}(i,k)\right|^4 \left|U(i,l)\right|^2 \Bigg]
			\end{aligned}
		}.
	\end{align}
	\hrulefill
	\vspace{-0.5cm}
\end{figure*}

\indent \emph{Proof:} See Appendix \ref{Appendix optimal filter}.

\indent By employing the optimal filter, the minimum MSE is attained in the vertex--fractional-frequency domain, resulting in optimal detection and filtering performance. The subsequent section presents experimental results that validate this optimal performance.

\indent \emph{Computational cost:} The proposed GFED domain filtering method (GFED-F) involves several key operations whose computational complexity we now analyze. For an input graph signal $\mathbf{x}$ of length $N$, the computation of the GFRFT, the GFED, and the filter transfer function matrix in the vertex–-fractional-frequency domain, each require a computational complexity of $\mathcal{O}(N^2)$. The computation of the eigendecomposition of the graph Laplacian $\mathbf{L}$, the GFT of GFED, and the estimated GFED through matrix multiplications each require a computational complexity of $\mathcal{O}(N^3)$. Therefore, the overall computational complexity of the proposed algorithm is dominated by these cubic-time operations and is $\mathcal{O}(N^3)$.

\indent Our proposed GFED-F operates in the vertex--fractional-frequency domain based on the GFED. Specifically, the filtering operation is conducted in the frequency--fractional-frequency domain by projecting the GFED onto the GFT basis, thereby enabling a structured and interpretable filtering process. The filter transfer function matrix $\widehat{\mathbf{H}}$ resides in the frequency--fractional-frequency domain. Moreover, the GFED-F method avoids dependence on the clean signal by leveraging the prior GFED $E^{a}_{\mathbf{x}}$.

\emph{The graph Wiener filtering method in \cite{Perraudin2017Stationary}:} The ideal graph Wiener filtering method aims to minimize the MSE objective
\begin{align}  \label{Wiener MSE}
	\mathbb{E}\left\{ \left\| \mathbf{H}_w \mathbf{y} - \mathbf{x} \right\|_2^2 \right\},
\end{align}
where $\mathbf{H}_w$ is a spatial domain filter. Under the assumption that the observation covariance $\mathbb{E}\left\{ \mathbf{y} \mathbf{y}^{\mathrm{H}} \right\}$ is invertible and that the original graph signal $\mathbf{x}$ is deterministic, the solution is given in closed form as
\begin{align}
	\mathbf{H}_w = \frac{\mathbf{x} \mathbf{x}^{\mathrm{H}}}{\sigma^2 + \|\mathbf{x}\|^2},
\end{align}
which requires access to the pure signal $\mathbf{x}$. This is a non-parametric filtering approach performed entirely in the vertex domain with a computational complexity of $\mathcal{O}(N^2)$.

\emph{The optimal GFRT domain filtering method (OGFRFT-F) \cite{Ozturk21}:} 
The OGFRFT-F minimizes the MSE objective
\begin{align}  \label{OGFRFT-F MSE}
	\mathbb{E}\left\{ \left\| \mathbf{F}_G^{-a} \mathbf{H}_o \mathbf{F}_G^a \mathbf{y} - \mathbf{x} \right\|_2^2 \right\},
\end{align}
where $\mathbf{H}_o$ is a diagonal matrix in the GFRFT domain. This method performs filtering directly in the spectral domain defined by the GFRFT basis and is a parametric spectral domain filtering method. In essence, OGFRFT-F can be regarded as a constrained version of graph Wiener filtering, where the filter is restricted to be diagonal in the GFRFT domain. OGFRFT-F has a computational complexity of $\mathcal{O}(N^4)$.

\emph{The GED-based Wiener filtering method (GED-WF) \cite{Yagan20}:} 
The GED-WF minimizes the MSE objective
\begin{align}  \label{GED-WF MSE}
	\mathbb{E}\left\{ \left\| \left( \mathbf{F}_G^{-1} \circ \mathbf{H}_{g} \right) \mathbf{F}_G \mathbf{y} - \mathbf{x} \right\|_2^2 \right\},
\end{align}
where $\mathbf{H}_{g}$ is a filter defined in the vertex--frequency domain. \eqref{GED-WF MSE} is mathematically equivalent to \eqref{Wiener MSE}. However, the graph Wiener filtering method requires access to the clean signal $\mathbf{x}$, which limits its practical applicability. To address this limitation, some algorithms are proposed in \cite{Yagan20} to estimate the filter $\mathbf{H}_g$ in the vertex--frequency domain without requiring knowledge of $\mathbf{x}$. These methods project the signal onto the GFT basis for filtering and maintain a computational complexity of $\mathcal{O}(N^3)$.

\indent In summary, both OGFRFT-F and GED-WF can be regarded as variants of the ideal graph Wiener filtering framework. In contrast, our proposed GFED-F method is fundamentally different in its formulation and mechanism, and thus also intrinsically distinct from OGFRFT-F and GED-WF. Table~\ref{comparison} summarizes the key differences among the four methods in terms of prior knowledge, projection basis, operational and filtering domains, and computational complexity.

\begin{table*}[!t]
	\caption{Comparison of the four filtering methods. }\label{comparison}
	\centering
	\resizebox{1.0\linewidth}{!}{
	\begin{threeparttable}
	\begin{tabular}{c|ccccc}
		\hline
		Method & Prior Knowledge & Projection Basis & Operational Domain & Filtering Domain & Computational Complexity \\ \hline
		GFED-F & $E^{a}_{\mathbf{x}}$, $\sigma$ & GFT basis & vertex--fractional-frequency domain & frequency--fractional-frequency domain & $\mathcal{O}(N^3)$ \\
		graph Wiener filtering method & $\mathbf{x}$, $\sigma$ & canonical basis & spatial domain & spatial domain  & $\mathcal{O}(N^2)$ \\
		OGFRFT-F & $\mathbf{xx}^{\rm H}$, $\sigma$ & GFRFT basis & spatial domain & fractional-frequency domain & $\mathcal{O}(N^4)$\\
		GED-WF & prior-free & GFT basis & spatial domain & vertex--frequency domain & $\mathcal{O}(N^3)$\\ \hline
	\end{tabular}
	\begin{tablenotes}
	\item [*]  $\mathbf{x}$ is deterministic.
	\end{tablenotes}
\end{threeparttable}
}
\vspace{-0.4cm}
\end{table*}

\section{Numerical experiments}   \label{sec7}
\indent In this section, we conduct numerical experiments to detect chirp signals from Examples 1–-2 using the proposed GFED-F, and apply this method to several real-world datasets, thereby validating the correctness of the theory.

\subsection{Detection of graph chirp signals}
\indent \emph{Example 3:} For the signal $\mathbf{x}_1$ defined in Example 1, corrupted by complex circular Gaussian noise with a standard deviation of $\sigma = 0.3$, the fractional order parameter is selected as the chirp rate of $\mathbf{x}_1$, i.e. $a = 0.7$.

\indent \emph{Example 4:} For the signal $\mathbf{x}_2$ defined in Example 2, corrupted by complex circular Gaussian noise with a standard deviation of $\sigma = 0.4$, the fractional order parameter is selected as the chirp rate of $\mathbf{x}_2$, i.e. $a = 0.6$.

\begin{figure}[!t]
	\centering
	\subfloat[]{\includegraphics[width=1.7in]{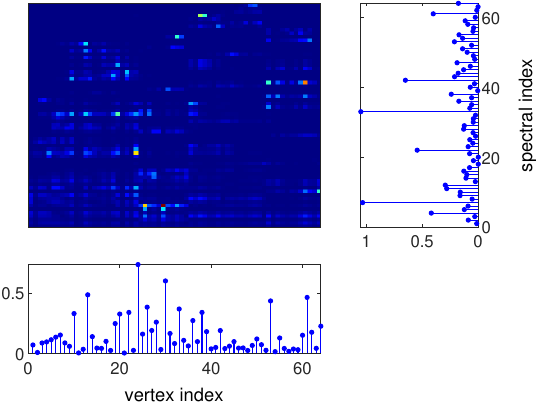}%
		\label{7a}}
	\hspace{\fill}
	\subfloat[]{\includegraphics[width=1.7in]{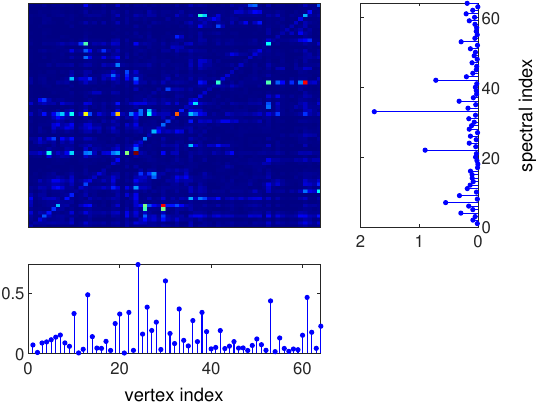}%
		\label{7b}}
	\
	\subfloat[]{\includegraphics[width=1.7in]{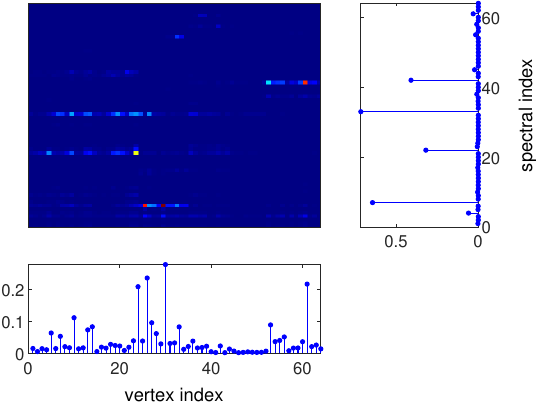}%
		\label{7c}}
	\hspace{\fill}
	\subfloat[]{\includegraphics[width=1.7in]{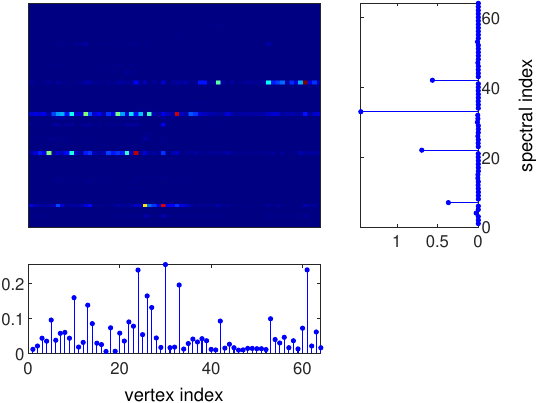}%
		\label{7d}} 
	\vspace{-0.1cm}
	\caption{Detection of graph chirp signals through filtering: Case of $\mathbf{x}_1$. (a) The GED of the noisy signal. (b) The GFED (a=0.7) of the noisy signal. (c) The filtered GED. (b) The filtered GFED (a=0.7).}
	\label{Fig7}
	\vspace{-0.5cm}
\end{figure}

\begin{figure}[!t]
	\centering
	\subfloat[]{\includegraphics[width=1.7in]{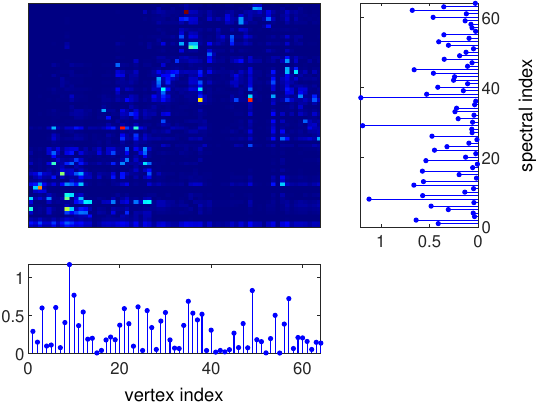}%
		\label{8a}}
	\hspace{\fill}
	\subfloat[]{\includegraphics[width=1.7in]{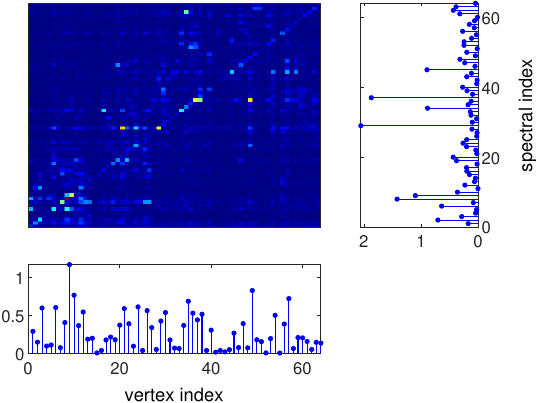}%
		\label{8b}}
	\
	\subfloat[]{\includegraphics[width=1.7in]{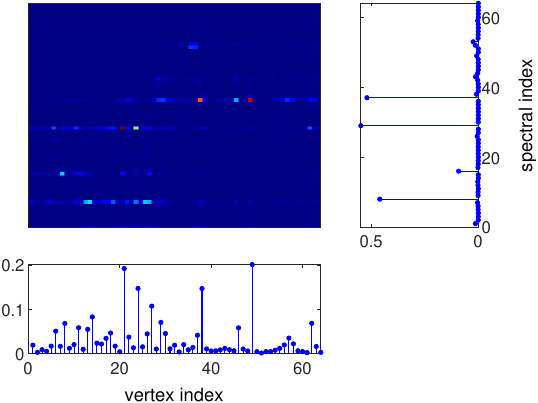}%
		\label{8c}}
	\hspace{\fill}
	\subfloat[]{\includegraphics[width=1.7in]{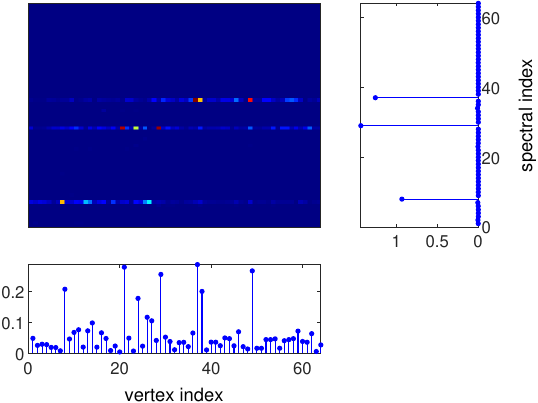}%
		\label{8d}} 
	\vspace{-0.1cm}
	\caption{Detection of graph chirp signals through filtering: Case of $\mathbf{x}_2$. (a) The GED of the noisy signal. (b) The GFED (a=0.6) of the noisy signal. (c) The GED of the filtered signal. (b) The GFED (a=0.6) of the filtered signal.}
	\label{Fig8}
	\vspace{-0.5cm}
\end{figure}

\indent Figs. \ref{Fig7}--\ref{Fig8} present the GEDs and GFEDs of the noisy signals in Examples 1--2, as well as those of the denoised versions obtained using the proposed filtering method. As observed, both representations of the noisy signals are significantly affected by noise, which obscures the vertex-–frequency structure and hinders the identification of key signal features. However, the vertex--fractional-frequency domain clearly exhibits enhanced robustness to noise after filtering, improving the detection of chirp signals. In contrast, the filtered GEDs still suffer from spurious spectral activations, which lead to analysis errors in the vertex–-frequency domain. As shown in Figs.~\ref{Fig7}\subref{7c} and \ref{Fig8}\subref{8c}, interference remains around vertex indices 30-–40 and frequency indices 50-–60. The corresponding Shannon entropy values of the filtered GEDs and GFEDs are listed in Table~\ref{Shannon-filtered}. The GFEDs consistently exhibit lower entropy values. These results confirm the effectiveness of the proposed filtering strategy in enhancing vertex--fractional-frequency representations for graph chirp signal detection.

\begin{table}[!t]
	\caption{Shannon entropy values of the filtered GEDs and GFEDWs. }\label{Shannon-filtered}
	\centering
	\begin{tabular}{c|c|c}
		\hline
		Signal & Distribution & Shannon entropy\\ \hline
		\multirow{2}{*}{$\mathbf{x}_1$} & GED & $45.0420$ \\ 
		& GFED ($a=0.7$) & $\textbf{44.4988}$\\ \hline
		\multirow{2}{*}{$\mathbf{x}_2$} & GED & $80.4329$ \\
		& GFED ($a=0.6$) & $\textbf{51.5346}$\\
		\hline
	\end{tabular}
\end{table}

\subsection{Denoising of real-data}
\indent We also conduct experiments on five real-world datasets: sea surface temperature (SST), particulate matter 2.5 (PM--25), the thickness data on the dendritic tree, the traffic volume data for Toronto, and the Minnesota road data. The first two datasets are discussed in \cite{Giraldo22}, while the latter three datasets can be accessed in \cite{graphsig}.

\indent For real-valued and positive signals, the vertex marginal distribution property of the GFED can be leveraged to restore the signal by performing denoising in the vertex domain. The SNR is used as the evaluation metric, calculated as $\mathrm{SNR}=20\mathrm{log}10\left(\frac{\left \|\mathbf{x}\right \| }{\left \| \mathbf{x}-\tilde{\mathbf{x}} \right \| }\right)$, where $\tilde{\mathbf{x}}$ is the filtered signal.

\indent To evaluate the effectiveness of the proposed GFED-F, we compare it with both graph spectral filtering methods and graph neural network (GNN)--based filtering methods. Specifically, the spectral filtering methods include OGFRFT-F \cite{Ozturk21} and GED-WF \cite{Yagan20}, while the GNN--based methods involve three classical architectures: Chebyshev graph convolutional network (ChebNet) \cite{Chebnet16}, graph attention network (GAT) \cite{GAT17}, and graph convolutional network (GCN) \cite{GCN17}. 

\indent \emph{Comparison with graph spectral filtering methods:} For the SST and PM--25 datasets, we use $k$-NN graph as in \cite{Giraldo22,Alik24} with $k=\{2,5,7\}$. Specifically, we use the SST data from months $T=\{50,120,270\}$, with zero-mean white Gaussian noise at noise levels $\sigma = \{15, 45, 60\}$. For the PM--25 data, we use data from days $T=\{50,120,270\}$, with zero-mean white Gaussian noise at noise levels $\sigma = \{15, 25, 35\}$. 

\begin{figure}[!t]
	\centering
	\subfloat[\label{Fig9a}]{\includegraphics[width=1.7in]{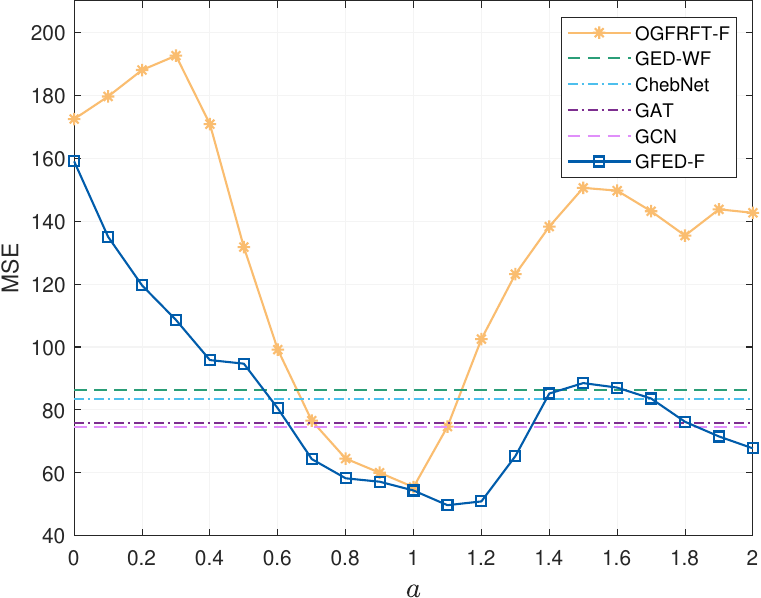}}
	\hspace{\fill}
	\subfloat[\label{Fig9b}]{\includegraphics[width=1.7in]{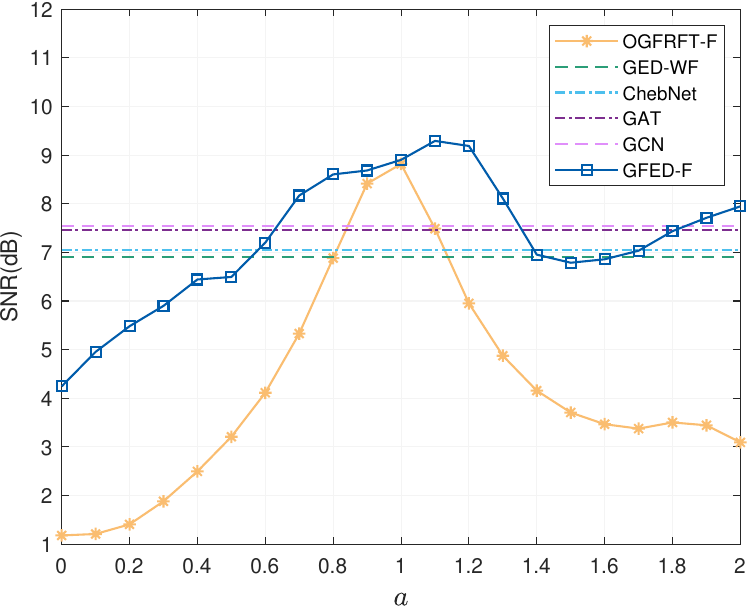}}
	\
	\subfloat[\label{Fig9c}]{\includegraphics[width=1.7in]{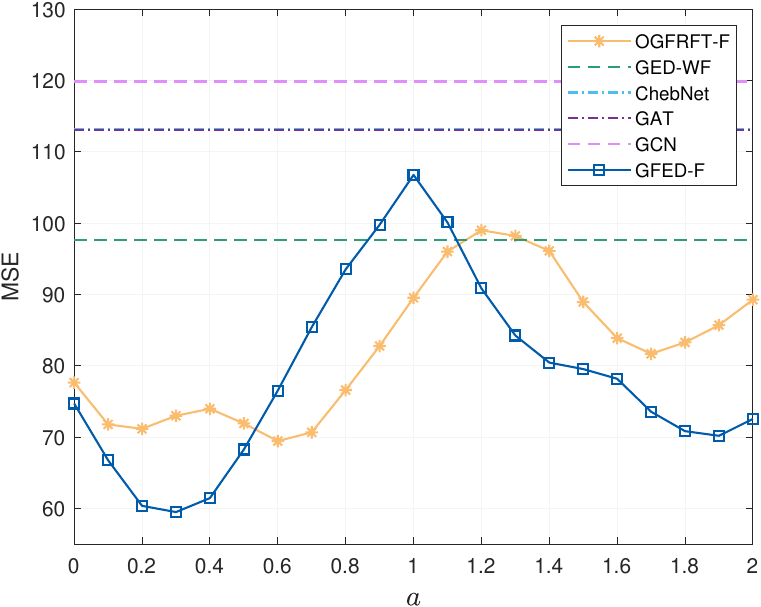}}
	\hspace{\fill}
	\subfloat[\label{Fig9d}]{\includegraphics[width=1.7in]{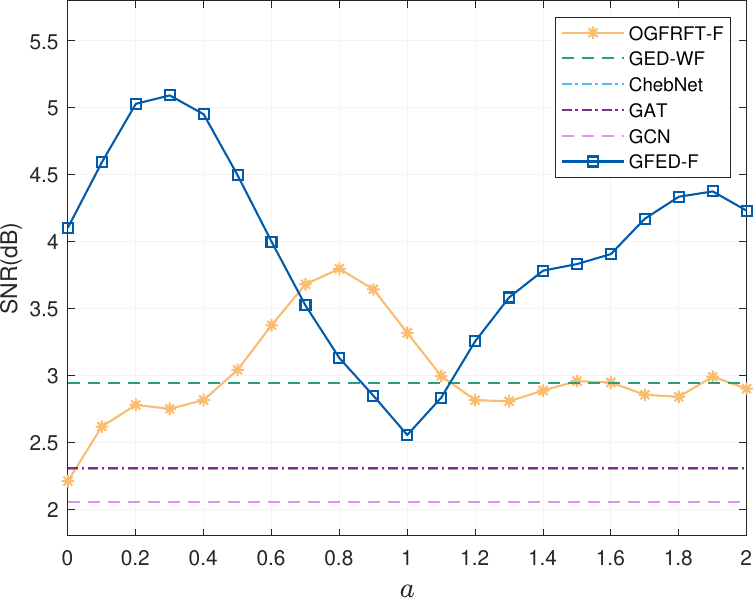}} 
	\vspace{-0.1cm}
	\caption{MSE and SNR comparison of six filtering methods on the SST dataset ($T=270$, $\sigma=40$) and the PM--25 dataset ($T=50$, $\sigma=25$). (a) MSE results on SST. (b) SNR results on SST. (c) MSE results on PM--25. (d) SNR results on PM--25.}
	\label{Fig9}
	\vspace{-0.5cm}
\end{figure}

\indent Fig. \ref{Fig9} presents the MSE and SNR line plots of the six filtering methods including OGFRFT-F, GED-WF, three GNN-based approaches (ChebNet, GAT, and GCN), and the proposed GFED-F. Figs. \ref{Fig9}\subref{Fig9a} and \ref{Fig9}\subref{Fig9b} depict the MSE and SNR results, respectively, for the SST dataset at $T=270$ with noise level $\sigma=40$. Figs. \ref{Fig9}\subref{Fig9c} and \ref{Fig9}\subref{Fig9d} depict the MSE and SNR results, respectively, for the PM--25 dataset at $T=50$ with noise level $\sigma=25$. The GFED-F consistently achieves strong performance across several fractional orders, and in many cases, it exceeds the best-case performance of OGFRFT-F. Notably, for certain fractional orders, GFED-F also outperforms GED-WF as well as all three GNN-based methods, further underscoring the effectiveness of the proposed framework for graph signal denoising.

\indent Table~\ref{table1} presents the MSE and SNR values for three spectral filtering methods: GED-WF, OGFRFT-F (optimized), and GFED-F (optimized), evaluated on the SST and PM--25 datasets under different graph connectivities ($k=2, 5, 7$) and noise levels. The results for OGFRFT-F and GFED-F are reported using their respective optimal fractional order parameters. Overall, GFED-F achieves the best performance in most settings, with lower MSE and higher SNR values compared to the other methods. In a few specific cases, such as the SST dataset at $T=50$ and $T=120$ under the $7$-NN graph with $\sigma=40$, and the PM--25 dataset at $T=50$ under the $2$-NN graph with $\sigma=15$, OGFRFT-F performs slightly better. Nevertheless, GFED-F consistently outperforms both OGFRFT-F and GED-WF in the majority of scenarios, demonstrating its robustness and effectiveness in graph signal denoising. 

\begin{table*}[!t]
	\caption{Comparison of MSE and SNR values on the SST and PM--25 datasets using OGFRFT-F (Optimized), GED-WF and GFED-F (Optimized).
	\label{table1}}
	\centering
	\resizebox{1.0\linewidth}{!}{
	\begin{tabular}{lcccccccccccc}
		\toprule[2pt]
		\multicolumn{2}{c}{\multirow{3}{*}{SST}} & \multicolumn{3}{c}{$T=50$} &&  \multicolumn{3}{c}{$T=120$} &&  \multicolumn{3}{c}{$T=270$} \\	
		\cmidrule{3-5} \cmidrule{7-9} \cmidrule{11-13}
		 ~ & ~ & $\sigma=15$	&  $\sigma=40$	& $\sigma=65$	&&  $\sigma=15$	&  $\sigma=40$	& $\sigma=65$	&& $\sigma=15$	&  $\sigma=40$	& $\sigma=65$ \\ 
		 \midrule
		~ & ~ & \multicolumn{11}{c}{$2$--NN} \\ 
		\midrule
		\multicolumn{1}{c}{\multirow{3}{*}{MSE}} &  OGFRFT-F &
		$49.0893$ ($a=1$) & $165.6258$ ($a=0.7$) & $178.4040$ ($a=1.9$) && $49.9294$ ($a=1$) & $160.4601$ ($a=0.7$) & $170.9464$ ($a=1.9$) && $51.4924$ ($a=1$) & $166.3954$ ($a=0.2$) & $178.5746$ ($a=1.9$) \\ 
		~& GED-WF & $92.4881$  & $202.0534$  & $513.8046$ && $91.2903$ &	$205.3636$ &	$515.3570$ && $92.0784$ &	$214.7092$ & $487.6849$ \\
		~& GFED-F & $\textbf{45.5943}$ ($a=1$) & $\textbf{106.9811}$ ($a=0.5$) & $\textbf{121.7892}$ ($a=1.7$) && $\textbf{46.0803}$ ($a=1$) & $\textbf{104.1300}$ ($a=0.5$) & $\textbf{117.0009}$ ($a=1.7$) && $\textbf{49.1533}$ ($a=1$) &  $\textbf{110.5426}$ ($a=0.4$) & $\textbf{123.8684}$ ($a=1.7$) \\ 
		\midrule
		\multicolumn{1}{c}{\multirow{3}{*}{SNR}} & OGFRFT-F & $9.3684$ ($a=1$) & $3.8365$ ($a=0.7$) & $2.5915$ ($a=0.6$) &&   
		$9.0481$ ($a=1$) & $3.6820$ ($a=0.7$) & $2.5120$ ($a=0.6$) && 
		$8.8801$ ($a=1$) & $3.4858$ ($a=0.7$) & $2.3535$ ($a=0.6$)  
		\\
		~& GED-WF &  $7.0569$ &	$3.6631$ & $-0.3902$ && $6.7542$ &	$3.2332$ & $-0.7627$ &&	$6.6161$ & $2.9392$ & $-0.6237$ \\
		~ & GFED-F & $\textbf{10.1287}$ ($a=1$) & $\textbf{6.4247}$ ($a=0.5$) & $\textbf{5.8617}$ ($a=1.7$) && 
		$\textbf{9.7233}$ ($a=1$) & $\textbf{6.1827}$ ($a=0.5$) & $\textbf{5.6765}$ ($a=1.7$) && 
		$\textbf{9.3421}$ ($a=1$) & $\textbf{5.8224}$ ($a=0.4$) & $\textbf{5.3281}$ ($a=1.7$)  \\
		\midrule
		~ & ~ & \multicolumn{11}{c}{$5$--NN} \\ 
		\midrule
		\multicolumn{1}{c}{\multirow{3}{*}{MSE}} &  OGFRFT-F & $33.2616$ ($a=1$) & $62.6326$ ($a=0.8$) & $75.8671$ ($a=0.8$) && $29.7670$ ($a=1$) & $56.0755$ ($a=1$) & $70.7181$ ($a=0.8$) && $27.2614$ ($a=1$) & $55.4753$ ($a=1$) & $74.6469$ ($a=0.8$) \\ 
		~& GED-WF & $35.0763$ &	$77.8422$ &	$120.0923$ &&	$34.0191$ &	$78.4322$ &	$122.4509$ 	&&	$35.9948$ &	$86.3369$ &	$135.7917 $ \\
		~& GFED-F & $\textbf{31.1213}$ ($a=1.1$) & $\textbf{56.9941}$ ($a=1.2$) & $\textbf{65.6439}$ ($a=0.7$) && $\textbf{27.2796}$ ($a=1.1$) & $\textbf{51.9049}$ ($a=1.1$) & $\textbf{64.0459}$ ($a=0.7$) && $\textbf{24.2385}$ ($a=1.1$) &  $\textbf{49.7348}$ ($a=1.1$) & $\textbf{71.6193}$ ($a=1.1$) \\ 
		\midrule
		\multicolumn{1}{c}{\multirow{3}{*}{SNR}} & OGFRFT-F & $11.4983$ ($a=1$) & $8.7418$ ($a=1$) & $7.4938$ ($a=1$) &&   
		$11.6211$ ($a=1$) & $8.8707$ ($a=1$) & $7.4377$ ($a=1$) && 
		$11.9022$ ($a=1$) & $8.8167$ ($a=1$) & $7.2472$ ($a=1$)  
		\\
		~& GED-WF & $11.2676$ &	$7.8056$ & $5.9226$ && $11.0412 $ &	$7.4135 $ & $5.4788$ &&	$10.6953$ &	$6.8957 $ & $4.9289$ \\ 
		~ & GFED-F & $\textbf{11.7872}$ ($a=1.1$) & $\textbf{9.1595}$ ($a=1.2$) & $\textbf{8.5458}$ ($a=0.7$) && 
		$\textbf{12.0000}$ ($a=1.1$) & $\textbf{9.2063}$ ($a=1.1$) & $\textbf{8.2935}$ ($a=0.7$) && 
		$\textbf{12.4126}$ ($a=1.1$) & $\textbf{9.2911}$ ($a=1.1$) & $\textbf{7.7074}$ ($a=1.1$) \\
		\midrule
		~ & ~ & \multicolumn{11}{c}{$7$--NN} \\ 
		\midrule
		\multicolumn{1}{c}{\multirow{3}{*}{MSE}} &  OGFRFT-F & $28.3506$ ($a=1$) & $\textbf{48.5869}$ ($a=0.8$) & $55.8849$ ($a=0.8$) && $25.8131$ ($a=1$) & $\textbf{45.7667}$ ($a=0.8$) & $53.2457$ ($a=0.8$) && $23.4273$ ($a=1$) & $49.7531$ ($a=0.8$) & $58.1912$ ($a=0.8$) \\ 
		~& GED-WF & $42.0824$ &	$77.7656$ &	$119.2170$ &&	$41.6953$ &	$78.3678$ &	$121.4248$ && $46.7268$ &	$85.0812$ &	$134.8790$ \\
		~& GFED-F & $\textbf{28.0820}$ ($a=0.9$) & $52.4520$ ($a=0.8$) & $\textbf{52.1758}$ ($a=0.7$) && $\textbf{25.3465}$ ($a=0.9$) & $47.1676$ ($a=0.8$) & $\textbf{48.8752}$ ($a=0.7$) && $\textbf{22.6316}$ ($a=1.1$) &  $\textbf{49.1505}$ ($a=0.8$) & $\textbf{55.5626}$ ($a=0.7$) \\ 
		\midrule
		\multicolumn{1}{c}{\multirow{3}{*}{SNR}} & OGFRFT-F & $12.1921$ ($a=1$) & $9.0967$ ($a=0.9$) & $7.9153$ ($a=0.9$) &&   
		$12.2400$ ($a=1$) & $9.2752$ ($a=0.9$) & $7.8787$ ($a=0.9$) && 
		$12.5604$ ($a=1$) & $9.1663$ ($a=0.9$) & $7.5997$ ($a=0.9$)  
		\\
		~& GED-WF & $10.4768$ &	$7.8099$ & $5.9544$ && $10.1575$ &	$7.4170 $ &	$5.5153$ && $9.5620$ & $6.9593$ & $4.9582$  \\
		~ & GFED-F & $\textbf{12.2335}$ ($a=0.9$) & $\textbf{9.5201}$ ($a=0.8$) & $\textbf{9.5431}$ ($a=0.7$) && 
		$\textbf{12.3192}$ ($a=0.9$) & $\textbf{9.6220}$ ($a=0.8$) & $\textbf{9.4675}$ ($a=0.7$) && 
		$\textbf{12.7105}$ ($a=1.1$) & $\textbf{9.3424}$ ($a=0.8$) & $\textbf{8.8099}$ ($a=0.7$)  
		 \\ \bottomrule[2pt]      
		 \toprule[2pt]
		 \multicolumn{2}{c}{\multirow{3}{*}{PM--25}} & \multicolumn{3}{c}{$T=50$} &&  \multicolumn{3}{c}{$T=120$} &&  \multicolumn{3}{c}{$T=270$} \\	
		 \cmidrule{3-5} \cmidrule{7-9} \cmidrule{11-13}
		 ~ & ~ & $\sigma=15$	&  $\sigma=25$	& $\sigma=35$	&&  $\sigma=15$	&  $\sigma=25$	& $\sigma=35$	&& $\sigma=15$	&  $\sigma=25$	& $\sigma=35$ \\ 
		 \midrule
		 ~ & ~ & \multicolumn{11}{c}{$2$--NN} \\ 
		 \midrule
		 \multicolumn{1}{c}{\multirow{3}{*}{MSE}} & OGFRFT-F & $\textbf{53.6247}$ ($a=0.1$) & $72.5014$ ($a=0.1$) & $86.6515$ ($a=0.1$) && 
		 $30.8939$ ($a=0.3$) & $39.0386$ ($a=0.4$) & $47.2836$ ($a=0.5$) && 
		 $14.6532$ ($a=0.9$) & $15.7741$ ($a=1$) & $16.2767$ ($a=1$) 
		 \\ 
		 ~& GED-WF & $120.6260$ &	$211.4029$ & $336.1629$ &&	$65.8810$ &	$154.9327$ & 	$260.6596$ && $54.0616$ & 	$121.8062$ & $218.4704$ \\
		 ~ & GFED-F & $56.2765$ ($a=0.3$) & $\textbf{64.5222}$ ($a=0.2$) & $\textbf{73.4911}$ ($a=0.1$) &&  
		 $\textbf{29.2637}$ ($a=0.4$) & $\textbf{33.8578}$ ($a=0.4$) & $\textbf{37.8861}$ ($a=0.5$) && 
		 $\textbf{13.0029}$ ($a=0.3$) & $\textbf{13.2016}$ ($a=0.2$) & $\textbf{13.9401}$ ($a=0.2$) 
		 \\ 
		 \midrule
		 \multicolumn{1}{c}{\multirow{3}{*}{SNR}} & OGFRFT-F & $\textbf{5.4916}$ ($a=0.9$) & $3.7791$ ($a=0.9$) & $2.7184$ ($a=0.9$) &&  
		 $4.1291$ ($a=2$) & $2.7812$ ($a=2$) & $2.2033$ ($a=0.9$) && 
		 $5.0868$ ($a=0.9$) & $4.3523$ ($a=1$) & $4.1595$ ($a=1$) 
		 \\
		 ~& GED-WF & $2.0236$ &	$-0.4131$ &	$-2.4275$ &&	$2.3473 $ & $-1.3665$ &	$-3.6258$ && $-0.5000 $ &	$-4.0278$ &	$-6.5650 $ \\
		 ~ & GFED-F & $5.3348$ ($a=0.3$) & $\textbf{4.7409}$ ($a=0.2$) & $\textbf{4.1757}$ ($a=0.1$) && 
		 $\textbf{5.8716}$ ($a=0.4$) & $\textbf{5.2383}$ ($a=0.4$) & $\textbf{4.7501}$ ($a=0.5$) && 
		 $\textbf{5.6885}$ ($a=0.3$) & $\textbf{5.6227}$ ($a=0.2$) & $\textbf{5.3863}$ ($a=0.2$) 
		 \\ \midrule
	
		 ~ & ~ & \multicolumn{11}{c}{$5$--NN} \\ 
		 \midrule
		 \multicolumn{1}{c}{\multirow{3}{*}{MSE}} & OGFRFT-F & $56.2678$ ($a=1.6$) & $69.4473$ ($a=0.6$) & $76.3163$ ($a=0.6$) && 
		 $32.1141$ ($a=0.8$) & $38.1410$ ($a=0.8$) & $40.8834$ ($a=0.8$) && 
		 $14.2646$ ($a=2$) & $16.5298$ ($a=0.7$) & $15.7226$ ($a=0.9$) 
		 \\ 
		 ~& GED-WF & $86.2320$ &	$97.6128$ &	$112.8108$ &&	$40.4191$ &	$50.6250$ &	$64.9516$ && $20.1283$ &	$32.2551$ &	$46.4879$ \\
		 ~ & GFED-F & $\textbf{51.2031}$ ($a=0.3$) & $\textbf{59.4986}$ ($a=0.3$) & $\textbf{62.6643}$ ($a=0.3$) &&  
		 $\textbf{27.1669}$ ($a=0.4$) & $\textbf{35.6684}$ ($a=0.5$) & $\textbf{37.9269}$ ($a=0.6$) && 
		 $\textbf{12.7965}$ ($a=1.7$) & $\textbf{14.0974}$ ($a=1.8$) & $\textbf{14.7062}$ ($a=0.2$) 
		 \\ 
		 \midrule
		 \multicolumn{1}{c}{\multirow{3}{*}{SNR}} & OGFRFT-F & $5.1140$ ($a=0.8$) & $3.7965$ ($a=0.8$) & $3.2534$ ($a=0.8$) &&  
		 $5.0900$ ($a=0.9$) & $4.2619$ ($a=0.9$) & $3.9119$ ($a=0.9$) && 
		 $4.4541$ ($a=0.9$) & $4.4498$ ($a=0.9$) & $5.1539$ ($a=0.9$) 
		 \\
		 ~& GED-WF & $3.4814$ &	$2.9430$ & $2.3145$ && $4.4691$ &	$3.4913 $ &	$2.4090$ && $3.7908 $ &	$1.7429 $ &	$0.1555$ \\
		 ~ & GFED-F & $\textbf{5.7451}$ ($a=0.3$) & $\textbf{5.0930}$ ($a=0.3$) & $\textbf{4.8678}$ ($a=0.3$) && 
		 $\textbf{6.1945}$ ($a=0.4$) & $\textbf{5.0121}$ ($a=0.5$) & $\textbf{4.7455}$ ($a=0.6$) && 
		 $\textbf{5.7580}$ ($a=1.7$) & $\textbf{5.3375}$ ($a=1.8$) & $\textbf{4.8040}$ ($a=0.2$) \\
		 \midrule
		 ~ & ~ & \multicolumn{11}{c}{$7$--NN} \\ 
		 \midrule
		 \multicolumn{1}{c}{\multirow{3}{*}{MSE}} & OGFRFT-F & $55.0139$ ($a=1.9$) & $76.2995$ ($a=0.1$) & $85.9867$ ($a=0.1$) && 
		 $35.5412$ ($a=0.9$) & $45.5639$ ($a=1$) & $50.5346$ ($a=1.1$) && 
		 $15.7331$ ($a=0.1$) & $16.0110$ ($a=1.3$) & $14.6749$ ($a=1.2$) 
		 \\ 
		 ~& GED-WF & $88.7646$ &	$103.0257$ & $118.5455$ &&	$41.0170$ &	$52.0031$ &	$66.8935$ && $20.2716$ &	$31.9256$ &	$46.6131$ \\ 
		 ~ & GFED-F & $\textbf{48.9340}$ ($a=0.3$) & $\textbf{61.3034}$ ($a=0.3$) & $\textbf{63.0833}$ ($a=0.3$) &&  
		 $\textbf{27.9296}$ ($a=0.5$) & $\textbf{33.2386}$ ($a=0.4$) & $\textbf{35.4350}$ ($a=1.6$) && 
		 $\textbf{12.2012}$ ($a=0.3$) & $\textbf{12.8751}$ ($a=0.3$) & $\textbf{13.0939}$ ($a=0.3$) 
		 \\ 
		 \midrule
		 \multicolumn{1}{c}{\multirow{3}{*}{SNR}} & OGFRFT-F & $5.3990$ ($a=1.9$) & $3.2787$ ($a=2$) & $2.7927$ ($a=1.2$) &&  
		 $5.0276$ ($a=0.9$) & $3.9487$ ($a=1$) & $3.4990$ ($a=1.1$) && 
		 $4.1618$ ($a=1.3$) & $4.3174$ ($a=1.2$) & $4.8639$ ($a=1.2$) 
		 \\
		 ~& GED-WF & $3.3556$ &	$2.7086$ & $2.0992$ && $4.4053$ &	$3.3746 $ &	$2.2811$ && $3.7600$ & $1.7875$ & $0.1438$ \\
		 ~ & GFED-F & $\textbf{5.9419}$ ($a=0.3$) & $\textbf{4.9632}$ ($a=0.3$) & $\textbf{4.8389}$ ($a=0.3$) && 
		 $\textbf{6.0743}$ ($a=0.5$) & $\textbf{5.3185}$ ($a=0.4$) & $\textbf{5.0406}$ ($a=1.6$) && 
		 $\textbf{5.9649}$ ($a=0.3$) & $\textbf{5.7314}$ ($a=0.3$) & $\textbf{5.6582}$ ($a=0.3$) 
		 \\ \bottomrule[2pt]
	\end{tabular}
}
\vspace{-0.5cm}
\end{table*}

\begin{figure}[!t]
	\centering
	\subfloat[]{\includegraphics[scale=0.33]{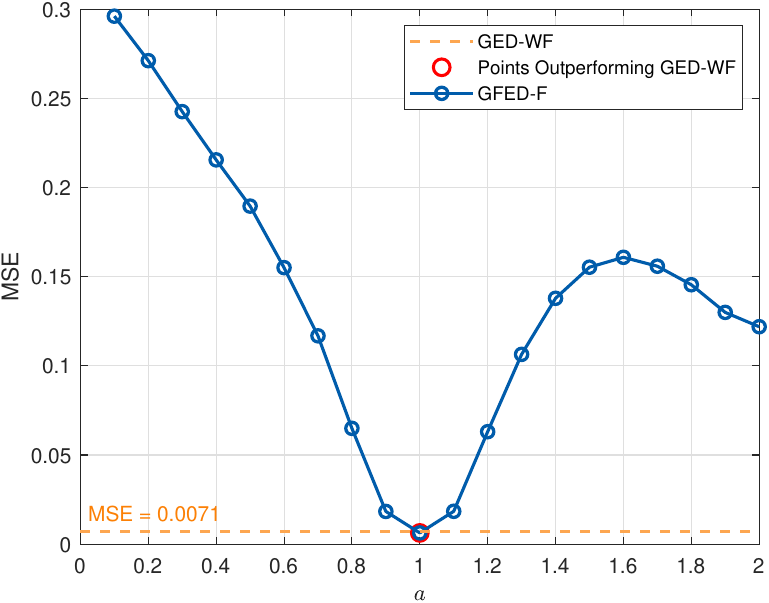}}
	\hspace{\fill}
	\subfloat[]{\includegraphics[scale=0.33]{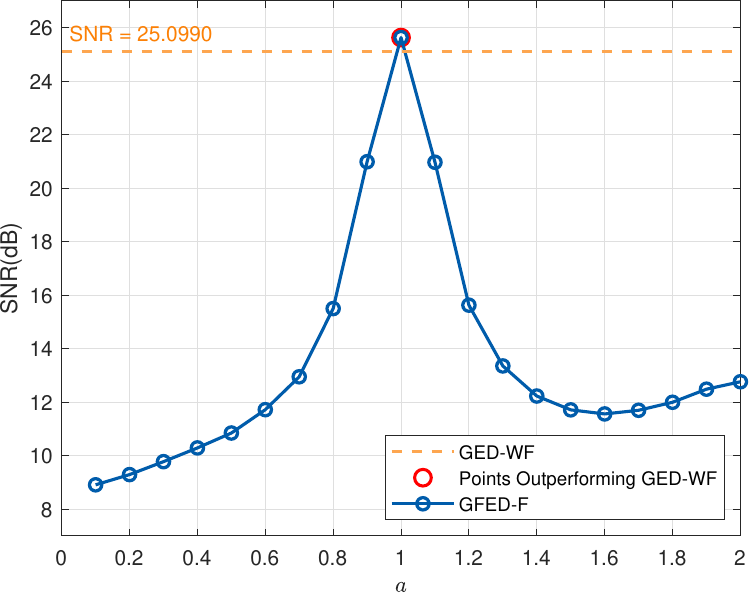}}
	\
	\subfloat[]{\includegraphics[scale=0.33]{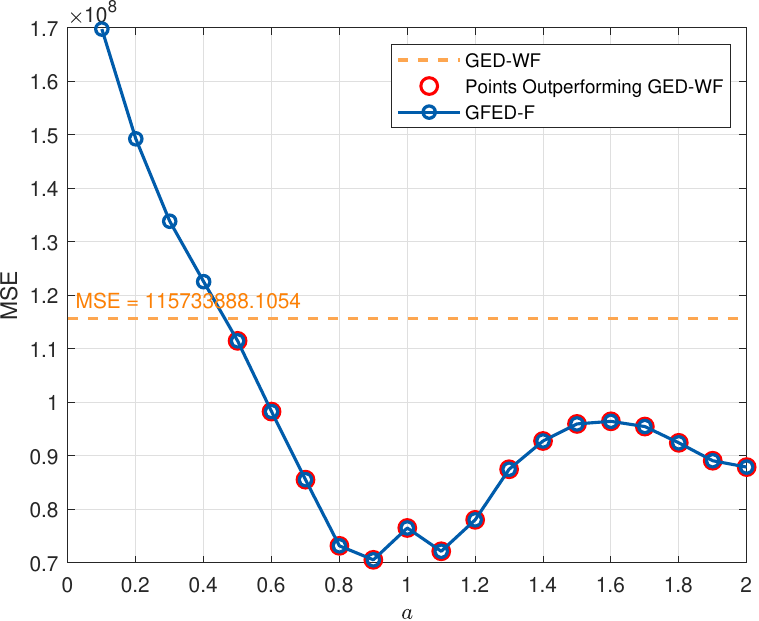}}
	\hspace{\fill}
	\subfloat[]{\includegraphics[scale=0.33]{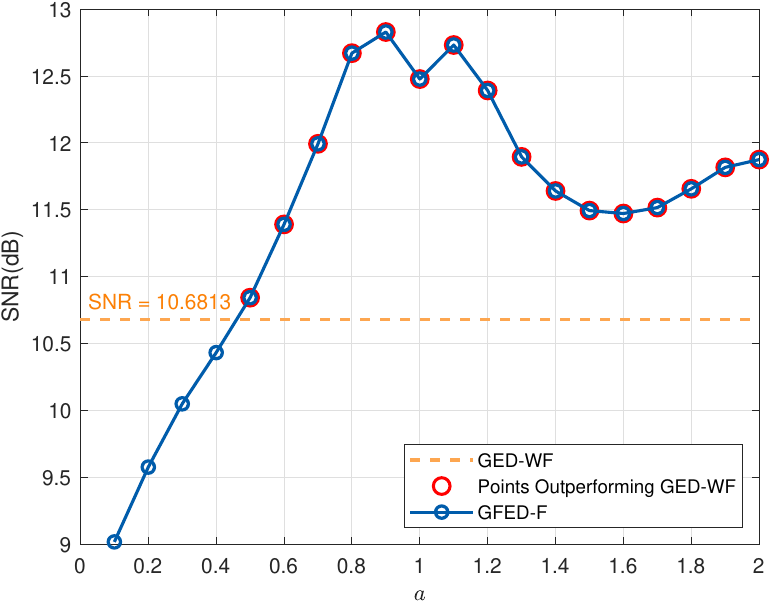}}
	\
	\subfloat[]{\includegraphics[scale=0.33]{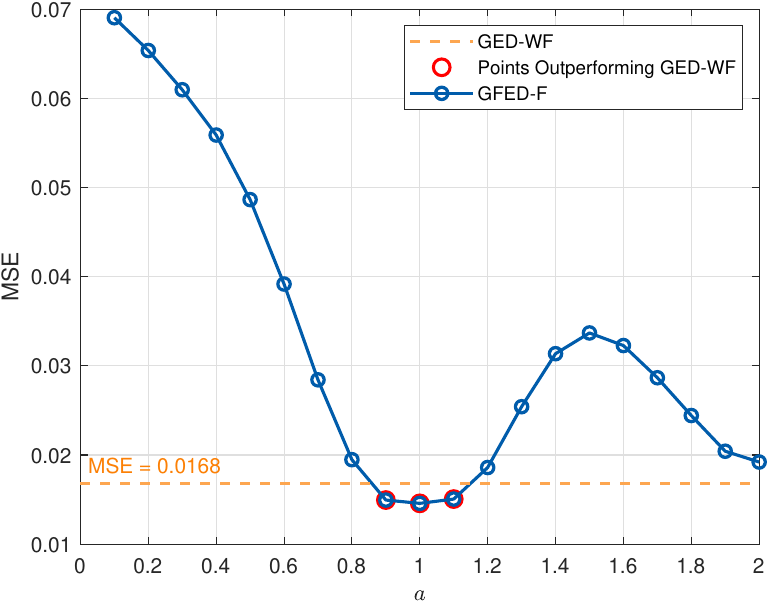}}
	\hspace{\fill}
	\subfloat[]{\includegraphics[scale=0.33]{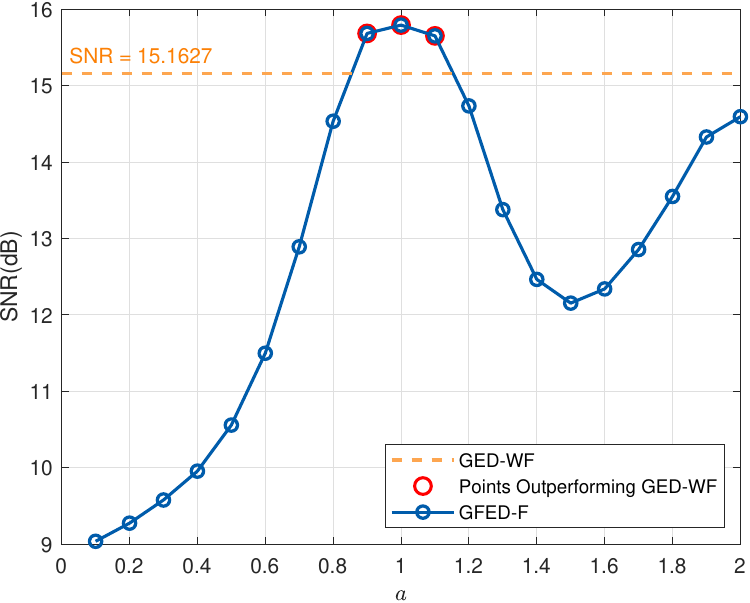}}
	\vspace{-0.1cm}
	\caption{Comparison of GED-WF and GFED-F. (a) MSE results on Dendritic tree. (b) SNR results on Dendritic tree. (c) MSE results on Toronto traffic volume. (d) SNR results on Toronto traffic volume. (e) MSE results on Minnesota road. (f) SNR results on Minnesota road. }
	\label{Fig10}
	\vspace{-0.2cm}
\end{figure}

\begin{figure}[!t]
	\centering
		\begin{minipage}[b]{0.33\linewidth}
			\centering
			\includegraphics[scale=0.21]{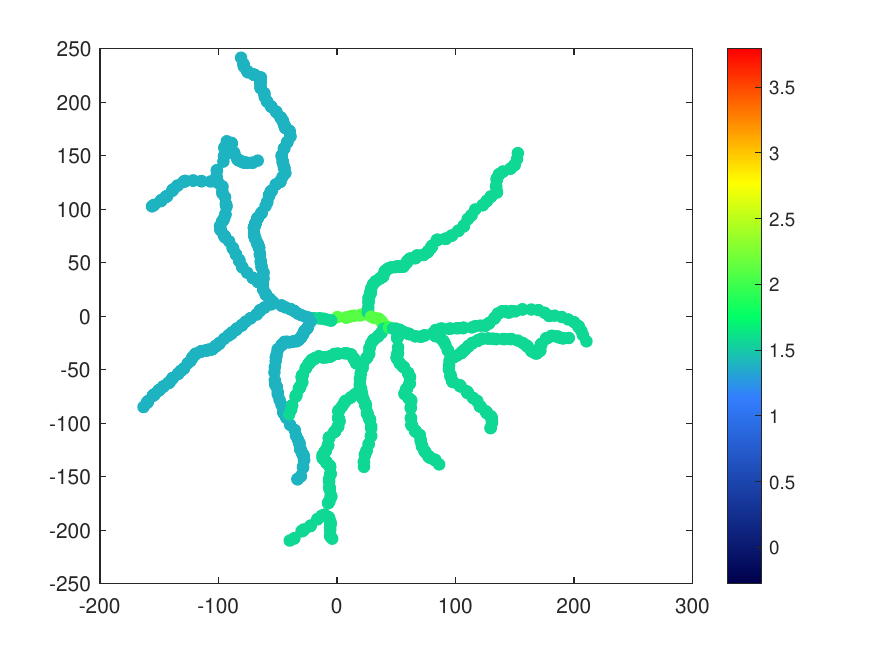}
			\includegraphics[scale=0.21]{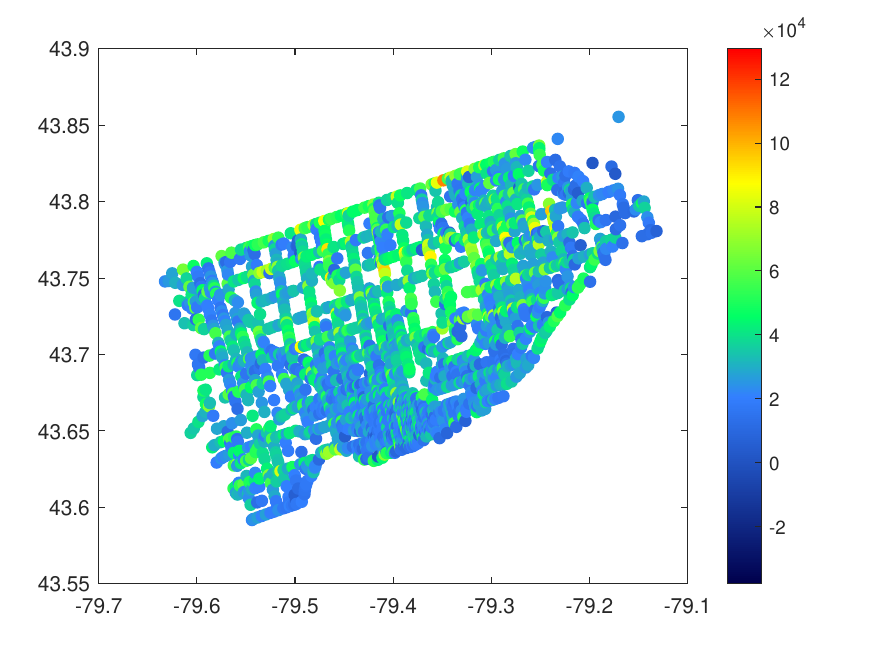}
			\includegraphics[scale=0.21]{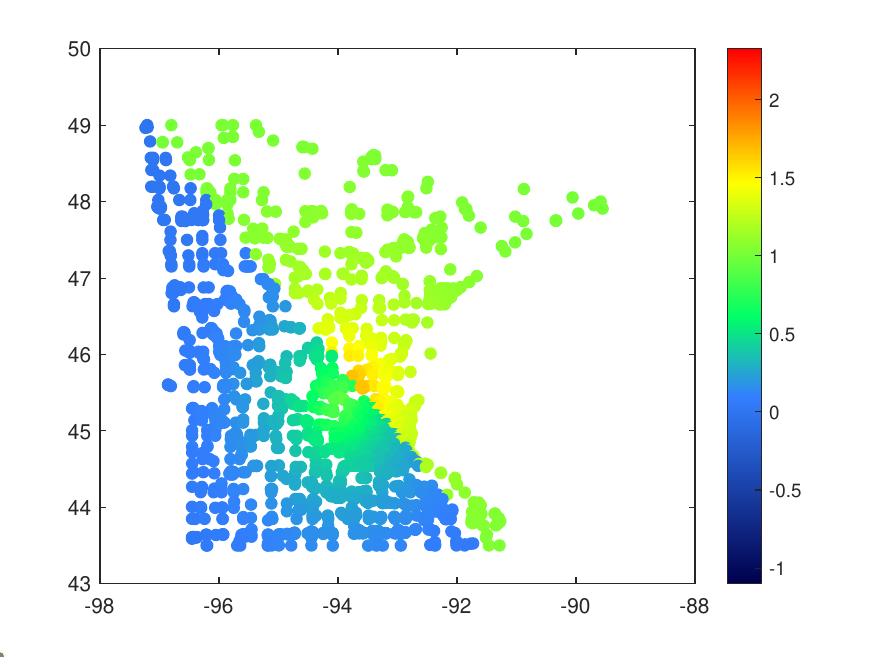}
			\\Original
		\end{minipage}%
		\begin{minipage}[b]{0.33\linewidth}
			\centering
			\includegraphics[scale=0.21]{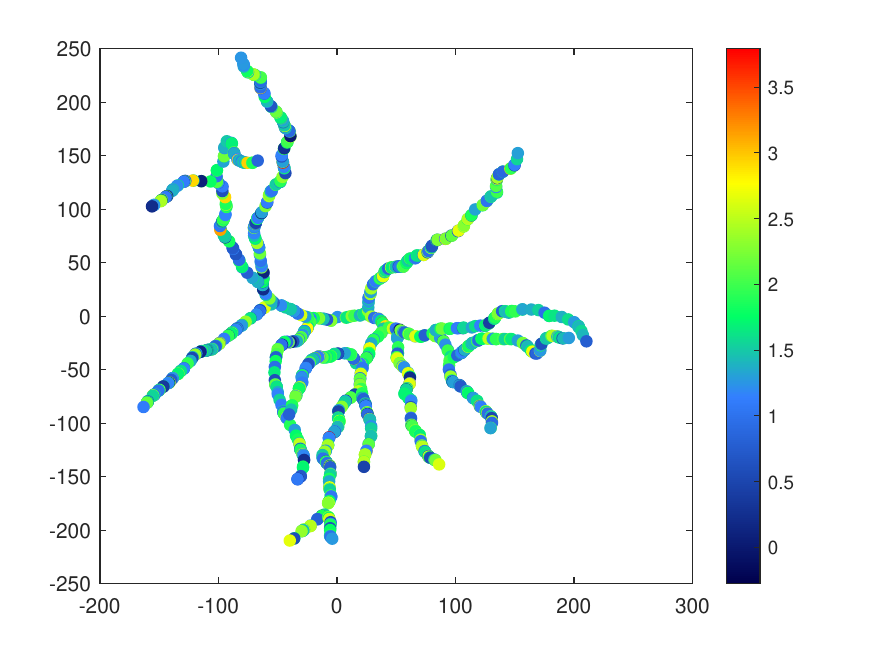}
			\includegraphics[scale=0.21]{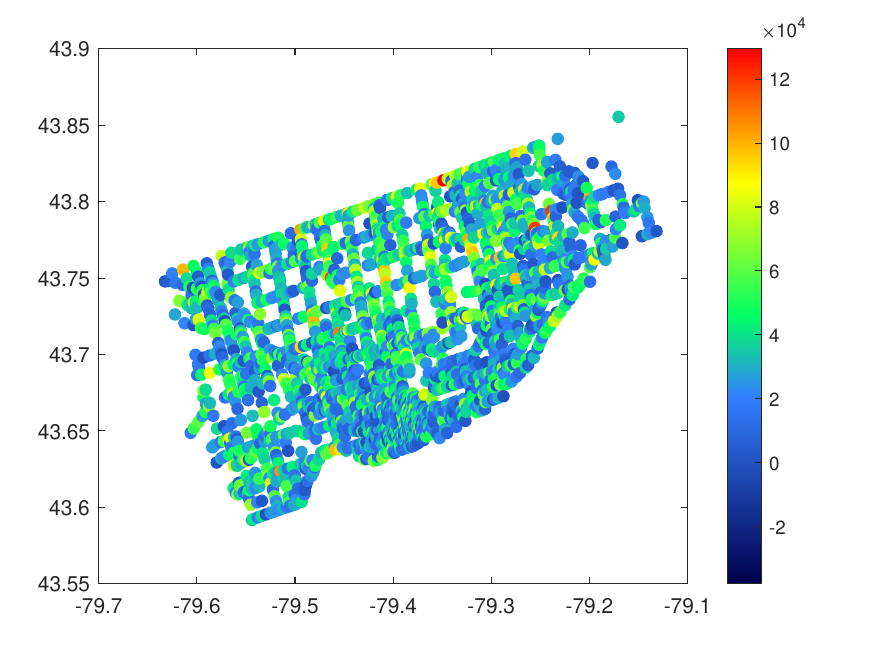}
			\includegraphics[scale=0.21]{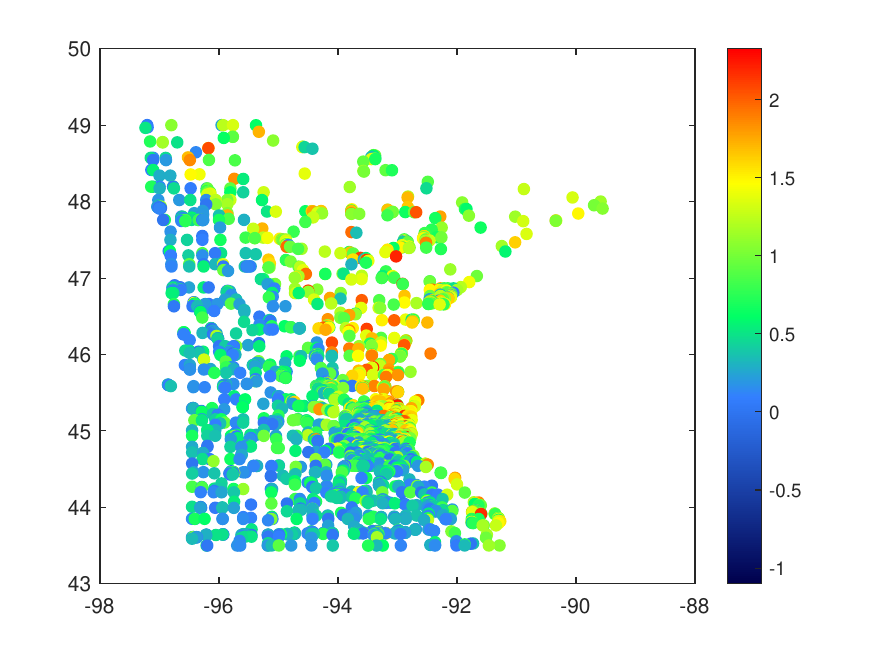}
			\\Noisy
		\end{minipage}%
		\begin{minipage}[b]{0.33\linewidth}
			\centering
			\includegraphics[scale=0.21]{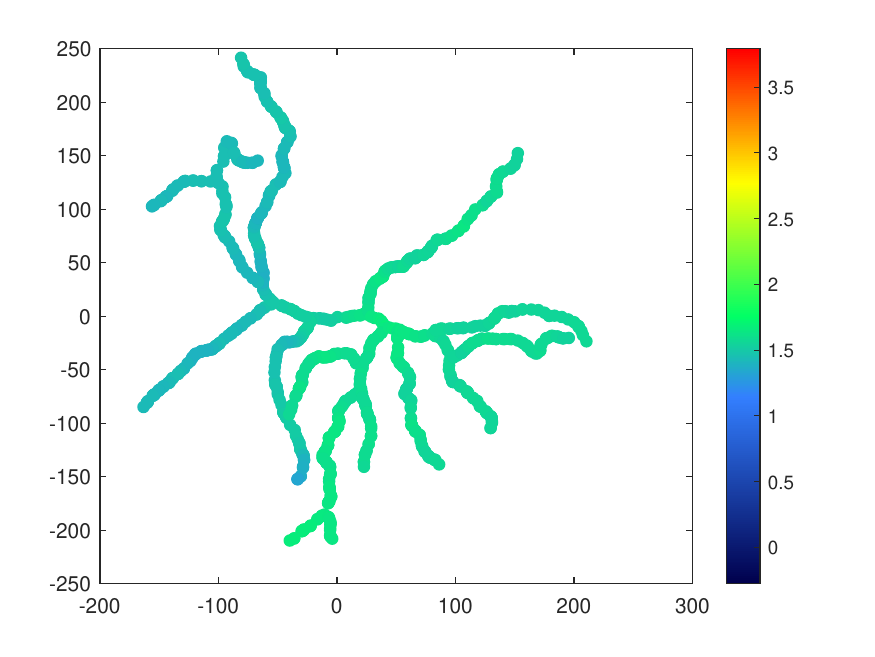}
			\includegraphics[scale=0.21]{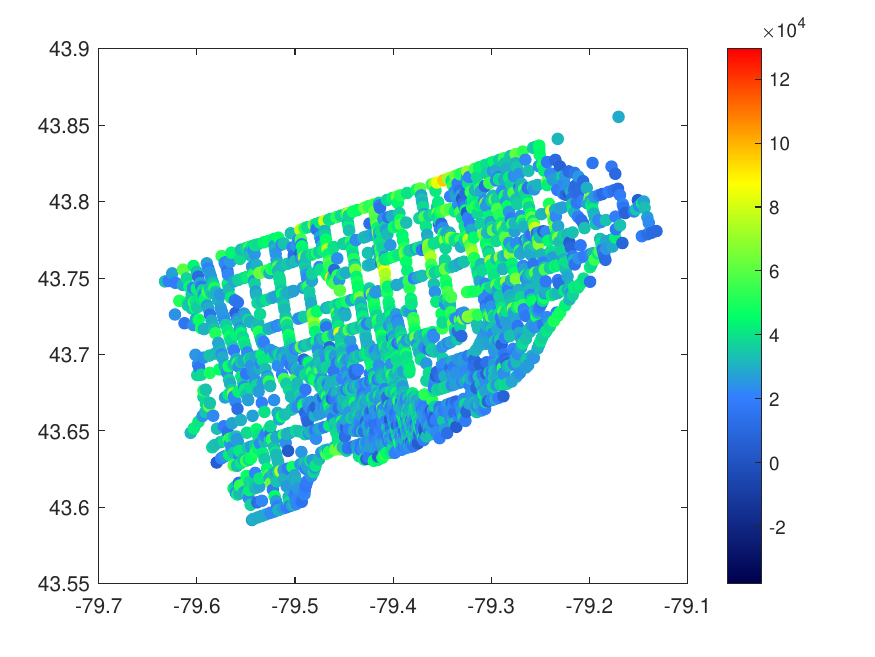}
			\includegraphics[scale=0.21]{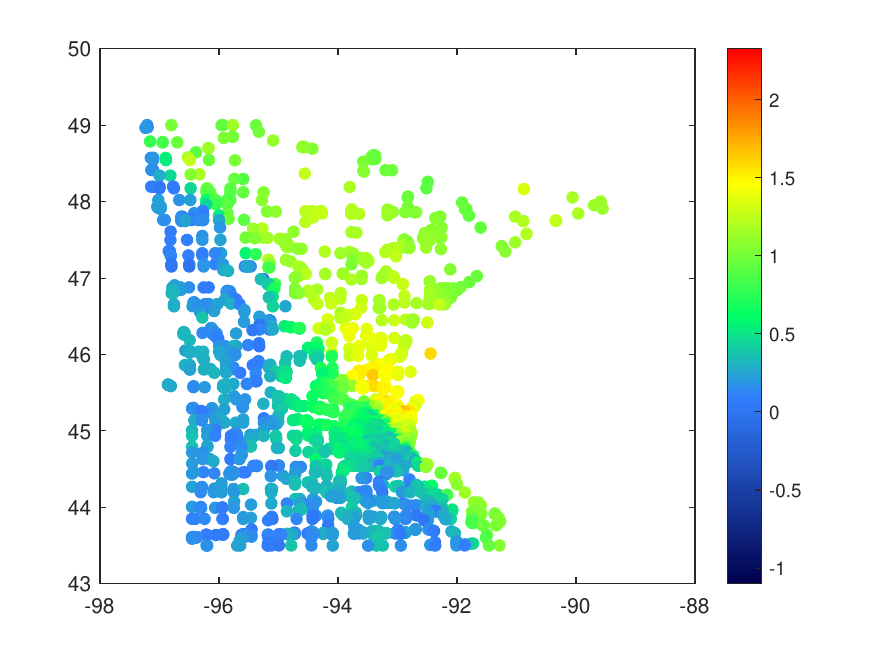}
			\\Filtered
		\end{minipage}%
	\vspace{-0.1cm}
	\caption{The original, noisy and filtered graphs of three datasets under optimal parameters: the thickness data on the dendritic tree (top row), the traffic volume data for Toronto (middle row), and the Minnesota road data (bottom row). }
	\label{Fig11}
	\vspace{-0.5cm}
\end{figure}

\indent For the thickness data on the dendritic tree, the traffic volume data from Toronto, and the Minnesota road data, we use the same noise levels as in \cite{Yagan20}, namely $8\mathrm{dB}$, $7\mathrm{dB}$, and $5\mathrm{dB}$,  respectively. Due to the relatively large scale of these datasets, the computational cost of OGFRFT-F is significantly higher. Therefore, we omit OGFRFT-F from this part of the comparison. Fig. \ref{Fig10} presents the MSE and SNR results comparing the proposed GFED-F with the GED-WF. It can be observed that for the dendritic tree data, our method achieves superior performance over GED-WF at the optimal fractional order. For the traffic volume and Minnesota road datasets, our method consistently outperforms GED-WF across multiple fractional orders. Fig.~\ref{Fig11} further illustrates the denoising performance of GFED-F under the optimal fractional order through visualizations on the three datasets. 

\indent \emph{Comparison with GNN--based filtering methods:} To evaluate the denoising performance of GFED-F against GNN--based graph filters, we compare it with three representative models: ChebNet, GCN, and GAT. For both the SST and PM2-5 datasets, we use the first 300 columns as input signals, excluding the frames at $T=50,120,270$, which are reserved for testing. The remaining frames are randomly split, with $80\%$ used for training and $20\%$ for validation. For the SST dataset, the learning rate is set to $0.001$, and the experiments are conducted under noise levels $\sigma=\{15, 45\}$. For the PM2-5 dataset, the learning rate is set to $0.0005$, with noise levels $\sigma=\{25, 35\}$. All neural network models are trained using the Adam optimizer (with a weight decay of $10^{-4}$). These settings ensure convergence for all models. Table \ref{table-GNN} presents the denoising performance of GFED-F compared to the three GNN--based methods in terms of MSE and SNR across both datasets and noise levels.

\begin{table}[!t]
	\caption{Comparison of MSE and SNR values on the SST and PM--25 datasets using ChebNet, GAT, GCN and GFED-F (Optimized). \label{table-GNN}}
	\centering
	\resizebox{1.0\linewidth}{!}{
		\begin{tabular}{lccccccccc}
			\toprule[2pt]
			\multicolumn{2}{c}{\multirow{3}{*}{SST}} & \multicolumn{2}{c}{$T=50$} &&  \multicolumn{2}{c}{$T=120$} &&  \multicolumn{2}{c}{$T=270$} \\	
			\cmidrule{3-4} \cmidrule{6-7} \cmidrule{9-10}
			~ & ~ & $\sigma=15$	&  $\sigma=40$	&&  $\sigma=15$	&  $\sigma=40$	&& $\sigma=15$	&  $\sigma=40$ \\ 
			\midrule
			\multicolumn{1}{c}{\multirow{4}{*}{MSE}} &  ChebNet & $49.8060$ &	$69.0983$  && $51.6344 $ &	$69.7478$ && $61.3659$ & 	$83.3395$ \\
			~ & GAT & $64.4516$ &	$68.5209$ &&
			$64.7811$ & $65.8596$ && $70.6245$ & $75.9799$ \\
			~ & GCN & $51.1914$	& $59.2552$ && $56.0345$ &	$60.8856$ &&	$67.7216$ &	$74.4438$ \\
			~& GFED-F & $\textbf{31.1213}$  & $\textbf{56.9941}$  && $\textbf{27.2796}$  & $\textbf{51.9049}$   && $\textbf{24.2385}$  &  $\textbf{49.7348}$\\ 
			\midrule
			\multicolumn{1}{c}{\multirow{4}{*}{SNR}} & ChebNet & $9.7449$ &	$8.3231$ && $9.2290 $ &	$7.9231$ && $8.3784$ &	$7.0492$ \\
			~ & GAT &  $8.6254$	& $8.3595$ && $8.2439$ &	$8.1722$ && $7.7681$ & $7.4507$ \\ 
			~ & GCN & $9.6258$ & $8.9905$ && $8.8739$ &	$8.5133$  &&	$7.9504 $ &	$7.5394$ \\
			~ & GFED-F & $\textbf{11.7872}$ & $\textbf{9.1595}$  && 
			$\textbf{12.0000}$ & $\textbf{9.2063}$  && 
			$\textbf{12.4126}$ & $\textbf{9.2911}$ 
			\\ \bottomrule[2pt]      
			\toprule[2pt]
			\multicolumn{2}{c}{\multirow{3}{*}{PM--25}} & \multicolumn{2}{c}{$T=50$} &&  \multicolumn{2}{c}{$T=120$} &&  \multicolumn{2}{c}{$T=270$} \\	
			\cmidrule{3-4} \cmidrule{6-7} \cmidrule{9-10}
			~ & ~ & $\sigma=25$	& $\sigma=35$	&&   $\sigma=25$	& $\sigma=35$	&& $\sigma=25$	& $\sigma=35$ \\ 
			\midrule
			\multicolumn{1}{c}{\multirow{4}{*}{MSE}} & ChebNet & $113.1868$ &$117.8153$ && $45.1324 $ &	$48.5219$ && $22.2523$ &	$21.7953$  \\
			~ & GAT & $113.0195 $ &	$114.9533$ && $47.2825$ &	$48.4982$ && $18.7667$ & $19.2397$  \\
			~ & GCN & $119.8559 $ &	$120.6042$ && $50.7381 $ &	$51.1548$  && $20.8578 $ &	$21.0155$ \\
			~ & GFED-F & $\textbf{59.4986}$& $\textbf{62.6643}$ &&  
			$\textbf{35.6684}$ & $\textbf{37.9269}$ && 
			$\textbf{14.0974}$ & $\textbf{14.7062}$ \\ 
			\midrule
			\multicolumn{1}{c}{\multirow{3}{*}{SNR}} & ChebNet &  $2.3001$ &	$2.1260$ && $3.9900 $ &	$3.6755$ && $3.3552$ & $3.4453$ \\
			~ & GAT & $2.3065$ &	$2.2328$ && $3.7879 $ &	$3.6777$ && $4.0950$ &	$3.9869$ \\
			~ & GCN & $2.0514$ & $2.0244$ && $3.4816$ &	$3.4461$ && $3.6362 $ &	$3.6035$ \\
			~ & GFED-F &  $\textbf{5.0930}$& $\textbf{4.8678}$ && 
			$\textbf{5.0121}$ & $\textbf{4.7455}$ && 
			$\textbf{5.3375}$ & $\textbf{4.8040}$ \\
			\bottomrule[2pt]
		\end{tabular}
	}
	\vspace{-0.5cm}
\end{table}

\section{Discussion} \label{sec8}
\indent A key observation from our study is that graph chirp signals are inherently dependent on the underlying graph topology. Different graph structures naturally give rise to different graph chirp signals. While the underlying graph is often constructed using conventional methods such as the $k$-NN graph, recent works have proposed data--driven approaches that learn the graph structure by optimizing the graph representation matrix \cite{NEURIPS2020_cdf6581c,wu2019graphwavenetdeepspatialtemporal,Li23Dynamic}. This insight leads to a novel perspective in which any real-world signal $\mathbf{x}$ can be interpreted as a graph chirp signal defined on a suitable graph structure. Given a signal $\mathbf{x} \in \mathbb{R}^N$, we first normalize it to obtain $\tilde{\mathbf{x}}$. Then, with a chosen graph chirp rate $a$ and initial graph frequency $k$, we extend $\tilde{\mathbf{x}}$ into a complete orthonormal basis of $\mathbb{R}^N$. This can be accomplished using various techniques such as Gram--Schmidt orthogonalization, Householder transformations, QR decomposition, or singular value decomposition. By inserting $\tilde{\mathbf{x}}$ as the $k$-th basis vector, we form the inverse GFRFT matrix 
\begin{align}
	(\mathbf{F}_{G}^{a})^{-1}=\left[\mathbf{u}_1^a,\cdots,\mathbf{u}_{k-1}^a,\tilde{\mathbf{x}},\mathbf{u}_{k+1}^a,\cdots,\mathbf{u}_{N}^a \right].
\end{align}
\indent The matrix
\begin{align}
	\mathbf{U} = \left( (\mathbf{F}_{G}^{a})^{-1}\right)^{\frac{1}{a}}=\left[\mathbf{u}_1^a,\cdots,\mathbf{u}_{k-1}^a,\tilde{\mathbf{x}},\mathbf{u}_{k+1}^a,\cdots,\mathbf{u}_{N}^a \right]^{\frac{1}{a}}
\end{align}
can then be treated as the eigenvector matrix of a graph shift operator. By assigning a suitable diagonal eigenvalue matrix $\mathbf{\Lambda}$ based on the desired task objective (e.g., smoothing, denoising, compression), we can construct a graph shift operator $\mathbf{Z = U\Lambda U}^{-1}$. This approach provides a practical and adaptable way to build graph representations directly from data, making it broadly applicable to real-world signal processing problems.

\indent Moreover, the proposed filtering framework can be extended to non--linear graph signal models. For example, it is worth exploring whether our approach can be adapted to frameworks such as the one proposed in \cite{Kroizer22Bayesian}. Future work may investigate how to incorporate fractional-frequency representations into such non-linear settings, possibly by learning graph chirp rates in a model--aware manner or integrating them with advanced inference schemes.

\section{Conclusion}  \label{sec9}
In this paper, we formally defined graph chirp signals, a previously undefined class of signals in GSP. We proposed the GFED to enhance the analysis of these signals, offering a more expressive representation of graph signals in the vertex--fractional-frequency domain. We explored the marginal distribution properties of GFED, further enhancing its theoretical foundation and interpretability. We also introduced the GFGD as a flexible and generalized vertex--fractional-frequency distribution, along with the reduced interference GFED, which effectively suppresses cross-terms interference and improves signal clarity. Furthermore, we proposed a graph chirp signal detection method based on GFED domain filtering and performed several experiments on graph chirp signals, demonstrating its robustness in detecting graph chirp signals under noisy conditions. Moreover, according to the marginal distribution properties of GFED, the proposed method can be applied to real-world data, showcasing its effectiveness in denoising tasks. The results confirm the effectiveness of our framework in both signal detection and noise reduction tasks, highlighting its potential for broader applications in graph signal analysis.


\appendices
\section{Proofs of Theorems 1--4} \label{Appendix Theorems 1--4}
\indent \emph{Proof of Theorem 1:} The GFRFT of the graph chirp signal $\mathbf{u}_{k}^{a}$ is 
\begin{align}
	\widehat{\left(\mathbf{u}_{k}^{a}\right)}_b = \mathbf{F}_{G}^{b} \left(\mathbf{F}_{G}^{a}\right)^{-1}\mathbf{e}_k = \mathbf{u}_{k}^{a-b}.
\end{align}
\indent When $a=b$, we have
\begin{align}
	\widehat{\left(\mathbf{u}_{k}^{a}\right)}_a = \mathbf{u}_{k}^{0} = \mathbf{e}_k.
\end{align}

\indent \emph{Proof of Theorem 2:} The $\ell_2$--norm of the graph chirp signal is
\begin{align}
	\|\mathbf{u}_k^{a}\|_2 = \mathbf{e}^{\rm T}_k \left(\left(\mathbf{F}_{G}^{a}\right)^{-1}\right)^{\rm H} \left(\mathbf{F}_{G}^{a}\right)^{-1}\mathbf{e}_k = \mathbf{e}^{\rm T}_k\mathbf{e}_k =1.
\end{align}

\indent \emph{Proof of Theorem 3:} The inner product of the graph chirp signals with different graph initial frequencies is 
\begin{align}
	\langle \mathbf{u}_k^{a}, \mathbf{u}_l^{a} \rangle = \mathbf{e}^{\rm T}_k \left(\left(\mathbf{F}_{G}^{a}\right)^{-1}\right)^{\rm H} \left(\mathbf{F}_{G}^{a}\right)^{-1}\mathbf{e}_l = \mathbf{e}^{\rm T}_k \mathbf{e}_l
	=\delta_{kl}.
\end{align}

\indent \emph{Proof of Theorem 4:} According to the GFRFT--invariance of graph chirp signals, the GFED of  $\mathbf{u}_{k_0}^{a_0}$ is given by
\begin{align} 
	E_{\mathbf{u}_{k_0}^{a_0}}^a(n,k) &= u_{k_0}^{a_0}(n)\,\overline{\widehat{\left( u_{k_0}^{a_0}\right)}_a(k)}\,\overline{u_{k}^{a}(n)} \nonumber\\
	&=u_{k_0}^{a_0}(n)\,\overline{ u_{k_0}^{a_0-a}(k)}\,\overline{u_{k}^{a}(n)} .
\end{align}
\indent In particular, when $a_0=a$, we have 
\begin{align}
	u_{k_0}^{a_0-a}(k) = e_{k_0}(k),
\end{align}
and therefore, we arrive the required result \eqref{Fractional energy concentration2}.

\section{Proof of Theorem 5} \label{Appendix filter transfer matrix}
\indent Inserting \eqref{GFED estimate} into \eqref{MSE}, we can obtain
\begin{align}
	&\mathbb{E}\left \{\left \| E^{a}_{\mathbf{x}}-E_{\widetilde{\mathbf{x}}}^{a} \right \|_{F}^{2}\right \}  \nonumber\\
	=& \mathbb{E}\left\{\sum_{n}\sum_{k}\left| E^{a}_{\mathbf{x}}(n,k)\right |^2  \right\} \nonumber\\
	&+ \mathbb{E}\left\{\sum_{n}\sum_{k}\left| \sum_{i}\widehat{E^{a}_{\mathbf{y}}}(i,k)\widehat{H}(i,k)U(n,i) \right|^2  \right\}  \nonumber\\
	&-\mathbb{E}\left\{\sum_{n}\sum_{k}\sum_{i}\overline{E^{a}_{\mathbf{x}}(n,k)}\widehat{E^{a}_{\mathbf{y}}}(i,k)\widehat{H}(i,k)U(n,i)  \right\} 
	\nonumber\\
	&-\mathbb{E}\left\{\sum_{n}\sum_{k}\sum_{i}E^{a}_{\mathbf{x}}(n,k)\overline{\widehat{E^{a}_{\mathbf{y}}}(i,k)}\thinspace\overline{\widehat{H}(i,k)}\overline{U(n,i)}  \right\} .
\end{align}
\indent Taking the derivative of the above equation with respect to the real and imaginary parts of $H(m,l)$, we obtain
\begin{align}
	&\mathbb{E}\left\{\sum_{i}\widehat{E^{a}_{\mathbf{y}}}(i,l)\overline{\widehat{{E^{a}_{\mathbf{y}}}}(i,l)}\widehat{H}(i,l)U(m,i)\right\} \nonumber\\
	=& \mathbb{E}\left\{\sum_{i}\widehat{E^{a}_{\mathbf{x}}}(i,l) \overline{\widehat{E^{a}_{\mathbf{y}}}(i,l)}U(m,i)   \right\},
\end{align}
which can be written as the matrix form as
\begin{align}
	\mathbf{U}\left(\mathbb{E}\left\{\widehat{E^{a}_{\mathbf{y}}}\circ\overline{\widehat{E^{a}_{\mathbf{y}}}}\right\}\circ\widehat{\mathbf{H}}\right)  =& \mathbf{U}\left(\widehat{E^{a}_{\mathbf{x}}}\circ\mathbb{E}\left\{\overline{\widehat{E^{a}_{\mathbf{y}}}}\right\}\right),
\end{align}
and therefore, we arrive the required result \eqref{filter transfer matrix}.

\section{Proof of Theorem 6}  \label{Appendix optimal filter}
\indent It is obvious that
\begin{align}
	\mathbb{E}\left\{E^{a}_{\mathbf{y}}(n,k)\right\}
	= E^{a}_{\mathbf{x}}(n,k) +\sigma^2 \left|U_{a}(n,k)\right|^2.
\end{align}
\indent Thus, we have
\begin{align}
	\mathbb{E}\left\{\widehat{E^{a}_{\mathbf{y}}}(l,k)\right\} 
	=& \mathbb{E}\left\{\sum_{i=1}^{N} E^{a}_{\mathbf{y}}(i,k)\overline{U(i,l)}\right\} \nonumber\\
	=& \widehat{E^{a}_{\mathbf{x}}}(l,k) + \sigma^2 \sum_{i}\left|U_{a}(i,k)\right|^2 \overline{U(i,l)}.
\end{align}

\indent And $\mathbb{E}\left\{\left|\widehat{E^{a}_{\mathbf{y}}}(l,k)\right|^2\right\} $ can be simplified as
\begin{align}
	&\mathbb{E}\left\{\left|\widehat{E^{a}_{\mathbf{y}}}(l,k)\right|^2\right\} \nonumber\\
	=&\sum_{i}\sum_{j}\mathbb{E}\left\{E^{a}_{\mathbf{y}}(i,k)\overline{E^{a}_{\mathbf{y}}(j,k)}\right\}\overline{U(i,l)}U(j,l).
\end{align}

\indent According to the property of zero-mean, complex circular Gaussian noise, we can obtain
\begin{align}  \label{eq35}
	&\mathbb{E}\left\{E^{a}_{\mathbf{y}}(i,k)\overline{E^{a}_{\mathbf{y}}(j,k)}\right\} \nonumber\\
	=& E^{a}_{\mathbf{x}}(i,k)\overline{E^{a}_{\mathbf{x}}(j,k)}
	+\sigma^2 x(i)\overline{\widehat{x}_{a}(k)}\thinspace\overline{U_{a}(i,k)}\left|U_{a}(j,k)\right|^2 \nonumber\\
	&+\sigma^2x(i)\overline{x(j)}\thinspace\overline{U_{a}(i,k)}U_{a}(j,k) \nonumber\\
	&+\sigma^2\left|\widehat{x}_{a}(k)\right|^2\overline{U_{a}(i,k)}U_{a}(j,k)\delta(i-j)  \nonumber\\
	&+\sigma^2\overline{x(j)}\widehat{x}_{a}(k)\left|U_{a}(i,k)\right|^2U_{a}(j,k) \nonumber\\
	&+\sum_{p}\sum_{q}\mathbb{E}\left\{w(i)\overline{w(j)}\thinspace\overline{w(p)}w(q)\right\} \overline{U_{a}(i,k)}U_{a}(j,k) \nonumber\\
	&\relphantom{=================} \times U_{a}(p,k)\overline{U_{a}(q,k)} .
\end{align}

\indent Since $\mathbb{E}\{|w(p)|^2w^*(i)w(j)\}=0$ for $i\neq j$, $\mathbb{E}\{|w(p)|^2w^*(i)w(j)\}=\mathbb{E}\{|w(p)|^2\}\mathbb{E}\{|w(i)|^2\}=\sigma^4$ for $i=j$, $\mathbb{E}\{|w(n)|^4\}=\mathbb{E}\left\{\left[w_R^2(n)+w_I^2(n)\right]^2\right\}=2\sigma^4$, we have
\begin{align} \label{eq36}
	&\mathbb{E}\left\{\left|\widehat{E^{a}_{\mathbf{y}}}(l,k)\right|^2\right\} \nonumber\\
	=& \left|\widehat{E^{a}_{\mathbf{x}}}(l,k)\right|^2  + \sigma^2\left|\sum_{i}x(i)\overline{U_{a}(i,k)}\thinspace\overline{U(i,l)} \right|^2 \nonumber\\
	& + \sigma^2 \sum_{i}\sum_{j}x(i)\overline{\widehat{x}_{a}(k)}\thinspace\overline{U_{a}(i,k)}\left|U_{a}(j,k)\right|^2 \overline{U(i,l)}U(j,l)  \nonumber\\
	&+ \sigma^2\sum_{i}\left|\widehat{x}_{a}(k)\right|^2\left|U_{a}(i,k)\right|^2\left|U(i,l)\right|^2  \nonumber\\
	& + \sigma^2\sum_{i}\sum_{j}\overline{x(j)}\widehat{x}_{a}(k)\left|U_{a}(i,k)\right|^2U_{a}(j,k)\overline{U(i,l)}U(j,l) \nonumber\\
	& + \sigma^4\left|\sum_{i}\left|U_{a}(i,k)\right|^2U(i,l)\right|^2  +\sigma^4\sum_{i}\left|U_{a}(i,k)\right|^2\left|U(i,l)\right|^2 \nonumber\\
	& 	+2\sigma^4\sum_{i}\left|U_{a}(i,k)\right|^4\left|U(i,l)\right|^2.
\end{align}

\indent With \eqref{filter transfer matrix}, \eqref{eq35}, and \eqref{eq36}, we have
\begin{align}
	\widehat{H}(l,k)  
	= \frac{\left|\widehat{E^{a}_{\mathbf{x}}}(l,k)\right|^2 + \sigma^2 \widehat{E^{a}_{\mathbf{x}}}(l,k) \sum_{i=1}^{N}\left|U_{a}(i,k)\right|^2 \overline{U(i,l)}}{\mathbb{E}\left\{\left|\widehat{E^{a}_{\mathbf{y}}}(l,k)\right|^2\right\}},
\end{align}
where $\mathbb{E}\left\{\left|\widehat{E^{a}_{\mathbf{y}}}(l,k)\right|^2\right\}$ is given by \eqref{eq36}.

\indent According to the inverse formula \eqref{inverse GFED  GFT}, we arrive the required result \eqref{optimal filter}.



\bibliography{mybib}
\bibliographystyle{IEEEtran}

\end{document}